\documentclass[fleqn,usenatbib]{mnras}

\usepackage{newtxtext,newtxmath}

\usepackage[T1]{fontenc}
\usepackage{ae,aecompl}
\usepackage[final]{pdfpages}

\usepackage{graphicx}	
\usepackage{amsmath}	
\usepackage{comment}    
\usepackage{float}
\usepackage[colorinlistoftodos]{todonotes}  %
\usepackage{caption}
\usepackage{subcaption}
\usepackage{epstopdf}

\title[AGN metallicity based on MaNGA]{Gas phase metallicity determinations  in nearby AGNs with SDSS-IV MaNGA: evidence of metal poor accretion}

\author[J. C. do Nascimento et al.]{Jana\'ina C. do Nascimento$^{1,4}$\thanks{E-mail:},
Oli L. Dors$^{1}$,
Thaisa Storchi-Bergmann$^{2,4}$,
\newauthor N\'icolas D. Mallmann$^{2,4}$, 
Rog\'erio Riffel$^{2,4}$,
Gabriele S. Ilha$^{3,4}$,
Rogemar A. Riffel$^{3,4}$,
\newauthor Sandro B. Rembold$^{3,4}$,
Alice Deconto-Machado$^{3,4}$,
Luiz N. da Costa$^{4,5}$,
Mark Armah $^{2,5}$
\\
$^{1}$Universidade do Vale do Para\'iba. Av. Shishima Hifumi, 2911, CEP: 12244-000, S\~ao Jos\'e dos Campos, SP, Brazil\\
$^{2}$Departamento de Astronomia, IF, Universidade Federal do Rio Grande do Sul, CP 15051, 91501-970, Porto Alegre, RS, Brazil\\
$^{3}$Departamento de F\'isica, CCNE, Universidade Federal de Santa Maria, 97105-900, Santa Maria, RS, Brazil\\
$^{4}$Laborat\'orio Interinstitucional de e-Astronomia - LIneA, Rua General Jos\'e Cristino 77, Rio de Janeiro, RJ - 20921-400, Brazil\\
$^{5}$Observat\'orio Nacional, Rua General Jos\'e Cristino, 77, Rio de Janeiro, RJ, 20921-400, Brazil\\
}
\date{Accepted XXX. Received YYY; in original form ZZZ}

\pubyear{2019}

\begin{document}
\label{firstpage}
\pagerange{\pageref{firstpage}--\pageref{lastpage}}
\maketitle

\begin{abstract} 
We derive the metallicity (traced by the O/H abundance) 
of the Narrow Line Region ( NLR) of 108 Seyfert galaxies as well as radial metallicity gradients along their galaxy disks and of these of a matched control sample of no active galaxies.
In view of that, observational data 
from the SDSS-IV MaNGA survey and strong emission-line calibrations taken from the literature were considered.
The metallicity obtained for the NLRs 
was compared to the value derived from the extrapolation of the radial oxygen abundance gradient, obtained from \ion{H}{ii} region estimates along the galaxy disk, to the  central part of the host galaxies. We find that, for most of the objects ($\sim 80\,\%$), the NLR metallicity is lower than the extrapolated value, with the average difference ($<D>$) between these estimates ranging from 0.16 to 0.30 dex. 
  We suggest that $<D>$ is due to 
the accretion of metal-poor gas to the AGN that feeds the nuclear supermassive black hole (SMBH), which is
drawn from a reservoir molecular and/or neutral hydrogen around the SMBH.
 Additionally,  we look for correlations between $D$ and the electron density ($N_{\rm e}$), [\ion{O}{iii}]$\lambda$5007 and  H$\alpha$ luminosities,  extinction coefficient ($A_{V})$ of the NLRs, as well as the stellar mass ($M_{*}$) of the host galaxies. Evidences of an  inverse correlation between the $D$ and the parameters $N_{\rm e}$, $M_{*}$ and $A_{\rm v}$ were found. 

\end{abstract}




\section{Introduction}

Emission line intensities present in the optical spectra of Active Galactic Nuclei (AGNs) and  star forming regions (SFs) are essential to estimate the physical properties of the gas phase, such as metallicity,
chemical abundances of  heavy elements (e.g. O, N, S),
electron temperature, and electron density of these objects.
In particular, due to their high luminosity and prominent emission lines, AGNs play a key role in studies of the chemical evolution of galaxies along the Hubble time.

Oxygen abundance is usually used to estimate gas-phase metallicity 
of  SFs 
(e.g. \citealt{kennicutt03, hagele08, yates12}) and of AGNs
(e.g. \citealt{SB98, Dors+15, dors20a, revalski18, flury20, dors21ep}).
This is due to  the oxygen is the third most abundant element after hydrogen and helium and it presents prominent optical emission lines ([\ion{O}{ii}]$\lambda$3726,$\lambda$3729, [\ion{O}{iii}]$\lambda$5007)
of its most abundant ions ($\rm O^{+}$, $\rm O^{2+}$) 
measured in most part of the spectra with high signal noise ratio.
In fact, \citet{skillman93} and \citet{dors20a} found that the
ion abundances with ionization stage higher $\rm O^{2+}$ do
not exceed $\sim5\,\%$ and $\sim20\,\%$ of the total O/H abundance
in SFs and AGNs, respectively. Therefore,  
such as in \citet{krabbe21},  hereafter we use metallicity ($Z$) and
oxygen abundance [in units of 12 + log(O/H)] interchangeably. 

The metallicity has been estimated 
preferably by comparisons between observational  emission line ratios
(e.g. [\ion{O}{ii}]($\lambda$3726+$\lambda$3729)/H$\beta$,
[\ion{O}{iii}]$\lambda$5007/H$\beta$, [\ion{N}{ii}]$\lambda$6584/H$\alpha$) with those predicted by photoionization models. One of the first $Z$ determinations in AGNs   was carried out by \citet{ferland83}, who compared observed and model predicted optical emission line ratios of low ionization nuclear emission-line regions (LINERs) and Seyferts using the {\sc Cloudy} code \citep[for the updated version see][]{Ferland+17}.
These authors found that the observed emission-line ratios were well reproduced by
photoionization models with metallicity in the range $0.1 \: \la \: (Z/{\rm Z_{\odot}}) \: \la 1.0$.
After this pioneering work, several studies have been performed with the goal to estimate $Z$ in AGNs
located at  low (e.g. \citealt{grazina84, ferland86,  SB98, groves06, Feltre+16, castro17, enrique19, carvalho19}) and high redshifts (e.g. \citealt{nagao06a, matsuoka09, matsuoka18,  nakajima18, dors18, dors19, mignoli19, guo20, ji20})
by using photoionization models. 
Particularly, \citet{SB98} proposed the first theoretical calibrations between
 strong optical emission-line ratios and metallicity, thus allowing estimations of $Z$ in large samples  of AGNs and the  development of new calibrations (e.g. \citealt{dors14, dors19,  castro17, carvalho19, dors21ep}). \citet{SB98} found that the $Z$ values of seven Seyfert~2 nuclei, derived from their calibrations,  are in agreement (within an uncertainty of $\sim$0.2 dex)
with those inferred from  extrapolation  of the oxygen abundance gradient to central parts of the hosting spiral galaxy. 
Thereafter, \citet{Dors+15} performed similar analysis but for a 
larger sample of AGNs (12 Seyfert~2)  and star-forming nuclei (33 objects).
\citet{Dors+15}   found  that  direct central oxygen abundances (i.e. estimates based on emission-line ratios from the nuclei) in some high-metallicity galaxies tend to be  lower than the extrapolated abundances. 

The existence of  lower oxygen abundances in the narrow line region of AGNs compared to the extrapolated value of abundance gradients is very  important in chemical evolution studies of galaxies, 
  since it indicates that the nuclear region could have distinct chemical evolution   as compared to the disk.
 This discrepancy can be due to the following physical processes:
\begin{itemize}
\item Spiral galaxies in the local universe can be accreting metal-poor gas from the outskirts of the disk onto the centres, a process more common at very high redshift 
(e.g. \citealt{cresci15, gillman21}). However, cases at low redhift have also been reported in dwarf galaxies, e.g. in  NGC 2915 \citep{Werk+10} and in NGC 4449 \citep{Kumari+17}.

\item The gas phase of AGNs can have a distinct  depletion of oxygen onto dust grains (e.g. \citealt{sternberg94, usero04}) in comparison to the one in  disk \ion{H}{ii} regions (of the order of 0.1 dex, e.g. \citealt{esteban98, meyer98, izotov06, jenkins09, whittet10}).
\item There is  metal-poor inflow of gas that can be originated from the capture of low mass companions to the nuclear region that ends up feeding the supermassive black hole (e.g. \citealt{thaisa07}).
\end{itemize}
\noindent To investigate which of these processes are acting in AGNs it is
necessary to estimate the metallicity in the entire galaxy, i.e. to consider estimates along the disk and in the  nuclear region. 
\citet{sanchez14}, by using observational data of galaxies 
($0.003 \: \la \: z\: \: \la 0.02$) selected by the Calar Alto Legacy Integral Field Area  (CALIFA) survey  \citep{sanchez12}, derived the oxygen gradients in about 300 spiral galaxies.   However,  the analysis carried out by \citet{sanchez14} was  mainly based on galaxies containing
star-forming nuclei. Another example is the CHAOS survey (\citealt{berg15, croxal15, croxal16, berg20, evan20,rogers21}), which has the
selection criterion to observe only galaxies with star-forming nuclei.
However, surveys such as Sydney Australian Astronomical Observatory Multi-object Integral Field Spectrograph (SAMI, \citealt{croom12}), although 
not dedicated to chemical evolution studies has provided
valuable information on the nature of AGNs and their host galaxies
(e.g. \citealt{comerford14, allen15, hampton17}).

In this work, we used data from the Mapping Nearby
Galaxies at Apache Point Observatory (MaNGA)  survey \citep{bundy15}.
We were motivated to consider these data due to the existence of  previous galaxy  selections by \citet{Rembold17}, which included a large sample of objects with
confirmed AGNs and their control sample containing non-active galaxies.
This advantage provides an excellent opportunity to compare
the direct AGN estimates with those obtained from oxygen extrapolation gradients, as they cover  larger parts of the disk in galaxies, including the nuclear region.
Additionally, it is useful to compare metallicity estimates in AGNs hosts sample with a sample of control galaxies, in order to verify if there is  any different effect on the metallicity gradient between them.
This paper is organized as follows: In Section~\ref{meth}, a brief description of the sample of objects
is presented as well as the methods used for estimating the
oxygen abundance  of the AGNs  and the oxygen abundance gradients of the sample. In Sect.~\ref{resdic}, we present the results and discussion. In Sect.~\ref{conc}, the summary and the conclusions are presented.
%

\section{Methodology}
\label{meth}
We selected from the MaNGA database  \citep{bundy15} a sample of spiral galaxies hosting AGNs. 
The emission-line intensities of the AGN and strong-line calibrations proposed in the literature were used to estimate
the oxygen abundances of these objects. In addition, emission-line intensities  of star-forming
regions located along the disk were used to calculate the oxygen abundance gradient
in each galaxy of the sample, and calibrations available in the literature were also considered. In the subsequent sections
a description of the methodology adopted to obtain the observational data and the abundance
values are presented.


\subsection{Observational data}
\label{obsc}

The sample of galaxies studied here comprise both galaxies hosting AGNs as well as a sample of control galaxies selected as in \citet{Rembold17}.  For each AGN was chosen two control non-active galaxies matching the AGN host stellar mass, morphology, distance and inclination. After the release of the MaNGA Product Launch 8 \citep[MPL-8, ][]{Aguado+19,Blanton+17,Wake+17,Law+15,Law+16,Yan+16,Drory+15,Gunn+06,Smee+13}, the number of observed AGNs with MaNGA has grown to 170 AGNs and 291 control galaxies as described in \citet{Riffel+21}.
We  classified  the nuclei (central region with 2.5 arcsec diameter) and the regions
along the disk of the galaxies according to their emission-line ratios in the diagnostic diagram [\ion{O}{iii}]$\lambda$5007/H$\alpha$ versus [\ion{N}{ii}]$\lambda$6584/H$\alpha$, proposed
by \citet{bpt81} (hereafter called BPT diagram) and on the WHAN diagram proposed by \citet{Cid+10}.

Firstly, to classify the regions in the sample as AGN-like and \ion{H}{ii}-like objects, 
we used the theoretical and the empirical criteria proposed by \citet{kewley01} and 
\citet{kauf03}, respectively. These criteria establish that regions with: 
\begin{equation}
\label{eq1}
\rm log([O\:III]\lambda5007/H\beta) \: > \: \frac{0.61}{log([N\:II]\lambda6584/H\alpha)-0.47}+1.19
\end{equation}
and
\begin{equation} 
\label{eq2}
\rm log([O\:III]\lambda5007/H\beta) \: < \: \frac{0.61}{log([N\:II]\lambda6584/H\alpha)-0.05}+1.3
\end{equation}
are classified as AGN-like objects, otherwise, as \ion{H}{ii}-like objects.
Additionally, we applied the criterion  proposed by  \citet{Cid+10}  to  separate 
AGN-like and  Low-ionization nuclear emission-line region (LINER) objects, given by
\begin{equation} 
\label{eq3}
\rm log([O\:III]\lambda5007/H\beta) \: > \: 0.47+log([N\:II]\lambda6584/H\alpha)\times1.01,
\end{equation}
where the values satisfying the above criterion correspond to Seyferts, otherwise, they are classified as LINERs.
Secondly, for each spaxel of the objects of our sample, the WHAN diagram proposed by \citet{Cid+10},  which takes into account the equivalent width of H$\alpha$ versus the  [\ion{N}{ii}]$\lambda$6584/H$\alpha$ line ratio, was considered to classify AGN-like and \ion{H}{II}-like objects.  This same procedure has been previously illustrated and applied in \citet{nascimento19}.

After applying the aforementioned  classification criteria, we selected the Seyfert type galaxies for our analysis using the final sample which consist of 98 Seyfert~2 and 10 Seyfert~1\footnote{The Seyfert~1 classification was obtained by visual inspection of all nuclear spectra, looking for the presence of broad components in H$\alpha$ and other Balmer lines.} galaxies as well as their respective control galaxies (145 in the total) for the present study. Among the 108 Seyfert galaxies, 71 are also classified as Seyfert using Sloan Digital Sky Survey (SDSS, \citealt{york00}) III/Baryonic Oscillation Spectroscopic
Survey (BOSS) collaboration  data and flux measurements from \citet{Thomas+13} -- the procedure originally used to identify the AGNs in the first paper of our series \citep{Rembold17}. The remaining 26 are classified as LINERs on the basis of SDSS-III data.
The larger number of Seyfert nuclei identified in the MaNGA data than those obtained using the 3.0 arcsec diameter SDSS-III fiber measurements, which is likely due to the better angular resolution of MaNGA. As gas surrounding the Seyfert region frequently shows LINER excitation \citep[e.g.][]{nascimento19}, the larger aperture of the SDSS-III data should be including more of the surrounding LINER region, as well as the possible contribution of emission from gas ionized by extra-nuclear evolved stars,
moving the  borderline objects from the Seyfert region to the LINER region of the BPT diagram.
The radial oxygen gradient in each object 
covers a region measured in effective radii ($R_{\rm e}$) ranging from $\sim$0.5 to $\sim$2.5 (see also \citealt{mingozzi20}).

The median point spread function (PSF) of the  MaNGA datacubes is estimated to have a full width at half maximum (FHWM) 
of about 2.5 arcsec \citep{Law+15}, which can be considered the resolution element in the data.  The redshift values of our sample are in the range
$0.02 \la z \la 0.08$. Assuming a spatially flat  cosmology with $H_{0}$\,=\,71 $ \rm km\:s^{-1} Mpc^{-1}$, $\Omega_{m}=0.270$, and $\Omega_{\rm vac}=0.730$  \citep{wright06}, 
each 
resolution element represents 
a region with diameters in the range $\sim$1-4 kpc
at the distance of the MaNGA sample. Thus, since \ion{H}{ii} regions have typical sizes of tens to a few hundreds parsecs,
each resolution element spectrum  comprises the flux of a complex of \ion{H}{ii} regions \citep{belfiore17}, 
and the physical properties derived 
represent an average value of these regions 
(e.g. \citealt{hagele08, krabbe14, rosa14}).
However, this fact has little influence on the radial oxygen abundance in spiral galaxies
because it is expected that \ion{H}{ii}  regions located at similar galactocentric
distances present about the same metallicity (e.g. \citealt{kennicutt03}) and similar
stellar content (e.g. \citealt{dors17a, igor19}).

Following \citet{Ilha19}, the emission-line fluxes from the MaNGA data cube were obtained by fitting a  single-Gaussian component using the Gas AND Absorption Line Fitting \citep[GANDALF; ][]{Marc06} routine. The choice of this routine was due to the fact that it fits both the stellar population spectra and the profiles of the emission lines, after the subtraction of the stellar population contribution. A detailed description of the fitting procedure is presented in \citet{Ilha19}.

\begin{figure*}
\includegraphics[width=1.8\columnwidth]{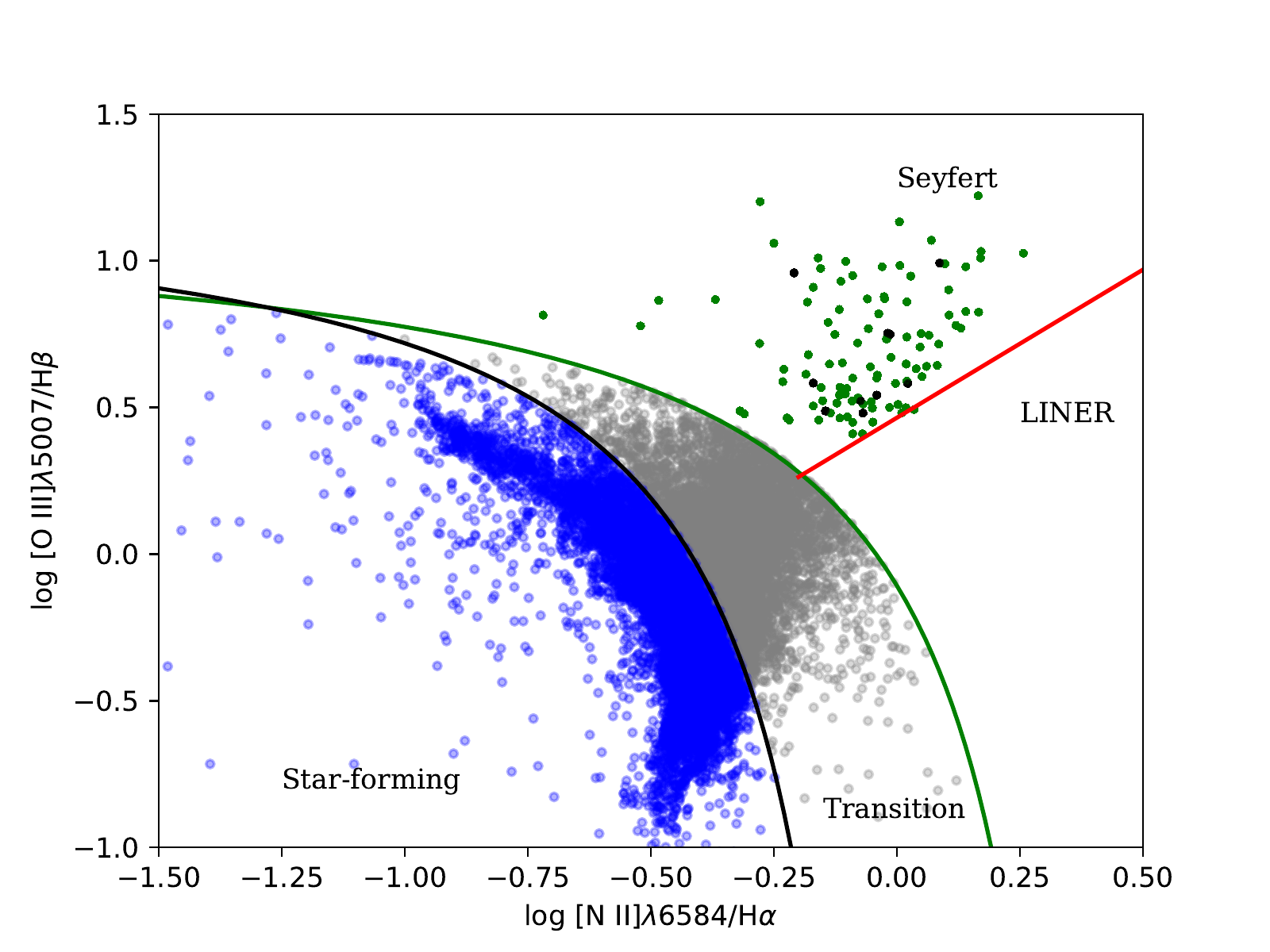}
\caption{[\ion{O}{iii}]$\lambda$5007/H$\beta$ versus  [\ion{N}{ii}]$\lambda$6584/H$\alpha$
diagnostic diagram for the AGN sample. The green (Seyfert~2) and the black points  (Seyfert~1) correspond to the emission-line ratios of the AGNs, obtained by summing the fluxes of all spaxels within an aperture of 2.5 arcsec in {diameter} in the central region of each galaxy. The blue and gray points are star-forming and transition regions, respectively, whose emission-line ratios were obtained from each spaxel of the datacubes.
Black and green lines correspond to the empirical and theoretical criteria
to separate AGN-like and \ion{H}{ii}-like objects proposed
by \citet{kauf03} and \citet{kewley01}, i.e. Eqs.~\ref{eq1} and \ref{eq2}, respectively. Red line represents the criterion  proposed by  \citet{Cid+10}  to  separate AGN-like and LINERs (Eq.~\ref{eq3}).}
\label{fig:bpt}
\end{figure*}

In Figure~\ref{fig:bpt} we present the BPT diagram for the central region of the AGN and all other spaxels for the AGN and control sample that correspond to \ion{H}{ii} and transition regions. The red line -- marking the separation between Seyfert and LINER excitation was obtained from \citet{kewley06}; the green line -- marking the separation between the transition and AGN regions  was obtained from \citet{kewley01} and the black line -- marking the separation between the starburst and transition regions, was obtained from \citet{kauf03}. Black circles and green triangles correspond to the emission-line ratios for the AGNs,  obtained by summing the fluxes of all spaxels within a central aperture of 2.5 arcsec diameter. 
The Seyfert~2 galaxies are represented by the black circles and Seyfert~1 by the green triangles.  Blue points are star-forming regions and grey points are transition regions. The diameter of the AGN nucleus of the galaxies is based on a fixed angular diameter of 2.5 arcsec  and
it ranges from $\sim$1 to $\sim$6 kpc according to the distance of each object. Thus, for some AGNs, an extended emission is observed.


 \subsection{Gas-phase metallicity determinations}

We estimated the gas phase metallicity in relation to the solar value ($Z/\rm Z_{\odot})$ for each AGN   and  the oxygen radial gradients  along the disk of each galaxy of the sample. 

In the AGN and in the disk \ion{H}{ii} region spectra of our sample emission lines sensitive to the electron temperature (e.g. [\ion{O}{iii}]$\lambda$4363, [\ion{N}{ii}]$\lambda$5755) were not detected and,
consequently, it was not possible to calculate the elemental abundances
by using the $T_{\rm e}$-method. 
Therefore, the 
O/H  abundance or the metallicity was calculated through calibrations  based on strong emission-lines. Several authors have investigated the O/H discrepancy values
obtained when distinct methods are assumed and  differences  up to 0.6 dex
have been found for \ion{H}{ii} regions (e.g. \citealt{Kewley-ellison+08, lopez12, pena12}). 
About the same discrepancy (i.e. up to 0.8 dex) is also derived for metallicity
AGN estimates taking into account different methods, being the highest discrepancy values derived for the low-metallicity regime ($\rm 12 + \log(O/H) \:  \la \: 8.5$, \citealt{dors20b}).

For  \ion{H}{ii} regions,  there seems to be a consensus that reliable calibrations are those
that produce O/H values similar (or near) to values derived through the 
$T_{\rm e}$-method (see \citealt{manuel17, enrique17}
for a review).  For AGNs, it is available in the literature
theoretical \citep{SB98}, semi-empirical  \citep{castro17, carvalho19, dors21} and empirical \citep{dors21ep}
calibrations between  strong optical narrow emission-lines and the metallicity or abundances. The recent empirical calibration for
AGNs proposed by \citet{dors21ep} requires measurements of the [\ion{O}{ii}]$\lambda$3727\footnote{[\ion{O}{ii}]$\lambda$3727 corresponds
to the sum of the $\lambda$3726 and $\lambda$3729 emission lines.} emission line, which is not available in our
data. Thus, it is not possible to use this calibration in our current study. 
It is beyond the scope of this work to investigate the O/H abundance discrepancy 
estimations in \ion{H}{ii} regions and AGNs derived when distinct 
methods are considered. However, it is worth to stress that the abundance values derived in this work can vary according to the calibrations assumed.
In what follows the calibrations used to estimate 
the  radial O/H abundance gradients  and the  AGN metallicity in our sample are presented.

\subsection{H\,II region calibrations}
\label{hiic}

Since the pioneering work of \citet{pagel79} several authors have proposed calibrations
between strong emission lines  and O/H abundance  for \ion{H}{ii} regions.
In particular, the first empirical calibration considering O/H abundances
derived through $T_{\rm e}$-method was the one proposed by \citet{SB94}, where 
the  $N2$=log([\ion{N}{ii}]$\lambda6584$/H$\alpha$) line ratio was considered
as metallicity indicator. After, \citet{leonid00, leonid01} inproved this methodology
taking into account the [\ion{O}{ii}]$\lambda3727$ and [\ion{O}{iii}]$\lambda5007$ emission lines.

In order to derive the oxygen abundance of \ion{H}{ii} regions along the galaxy disk of our sample and for the control sample (in the nuclei and in the disk \ion{H}{ii} regions), we considered the  assumption that calibrations based on O/H calculated via $T_{\rm e}$-method  (empirical calibrations) more reliable in comparison with theoretical calibrations. In this sense, we assumed the following empirical calibrations proposed
by \citet{Perez+09}: 
\begin{equation}
\mathrm {12 + \log(O/H)} = \mathrm{8.74 - 0.31} \times O3N2, 
\label{abund_HII}
\end{equation}
where
\begin{equation}
O3N2 =  \log \left(\frac{\rm{([\ion{O}{iii}]\,\lambda 5007})}{\rm{(H{\beta})}}\times \frac{\rm{ (H\alpha})}{\rm{([\ion{N}{ii}]\,\lambda6584)}}\right).
\label{eq_O3N2}
\end{equation}
This relation is valid  for $\rm 12+\log(O/H) \: \ga \: 8.0$. The $O3N2$ index was introduced by \citet{Alloin+79} as metallicity indicator. The other calibration considered in this work is based on the $N2$ parameter:
\begin{equation}
\mathrm {12 + \log(O/H)} = 9.07 + 0.79 \times N2. 
\label{abund_HII_PMC09}
\end{equation}

The distribution of O/H values along the disk of the Seyfert~2 galaxy  MaNGA ID 1-210646 as well as for its two control galaxies are shown in the second column of Fig.~\ref{fig_2}.  Analogous figures to Fig.~\ref{fig_2} are available as supplementary material for the remaining galaxies. The O/H values obtained from the  above relations (i.e. Eqs.~\ref{abund_HII} and \ref{abund_HII_PMC09}) for all  the spaxels 
 were used to obtain radial abundance gradients along the disk of each galaxy, 
 following the methodology proposed by \citet{riffel+21}.   Thus, we have  obtained mean azymuthal values and standard deviations for O/H in radial bins of 2.5 arcsec along the galaxy disks.  Thereafter, we used the following relation to fit the radial abundance distributions 
\begin{equation}
\label{graf}
Y=Y_{0}\: + grad \: Y \: \times \:R,    
\end{equation}
where $Y$ is a given oxygen abundance [in units of 12+log(O/H)], $R$ is the galactocentric distance (in units of arcsec),  $Y_{0}$ is the extrapolated value of the gradient to the galactic centre
($R=0$), and  $grad \: Y$ is the slope of the
distribution (in units of dex/arcsec). The third column in Fig.~\ref{fig_2} shows these gradients for the Seyfert~2 galaxy with MaNGA ID 1-210646 and its control galaxies.

\begin{table*}
\footnotesize
\caption{Parameters for the 107 Seyfert galaxies of our sample of AGNs in MaNGA-MPL8:
Columns: (1) Galaxy identification in the MaNGA survey. (2)-(3) slope ($grad \: Y$, in units of dex/arcsec) of the oxygen radial gradient  and the extrapolated value of the gradient to the galactic centre ($R=0$) considering
the fit of the Eq.~\ref{graf} to the abundance estimates along the entire disk of each object
and the (O/H)-$N2$ calibration (Eq.~\ref{abund_HII_PMC09}). (4)-(5) Same than for  the columns (2)-(3) but for estimations from (O/H)-$O3N2$ calibration (Eq.~\ref{abund_HII_PMC09}). (6)-(7) Estimates of $\rm 12 + \log(O/H)$ for the nuclear region of each galaxy obtained through
the calibrations proposed by \citet{carvalho19} and  \citet{SB98}, respectively.}
\label{table1}
\begin{tabular}{lcccccccc}
\hline
\noalign{\smallskip}
      
         &       \multicolumn{2}{c}{$N2$}          &    & \multicolumn{2}{c}{$O3N2$}&      &            \multicolumn{2}{c}{12+log(O/H)}       \\
     ID  &     slope               & (O/H)$_{0}$   & 	&  slope  & (O/H)$_{0}$	    &	   &	      ${\rm O/H})_{C}$  & $\rm (O/H)_{SB98f}$ \\
\cline{2-3}
\cline{5-6}
\cline{8-9}
\noalign{\smallskip}
(1)      &     (2)                 &     (3)       &    &  (4)    &      (5)        &      &             (6)            &       (7)       \\ 
1-44303  &   $-$0.004$\pm0.001$    & 8.80$\pm0.018$ & 	& $-$0.007$\pm0.002$   &	8.78$\pm0.017$   &	   &	       8.65	&    8.58 \\
1-460812 &   $-$0.004$\pm0.005$    & 8.79$\pm0.087$ & 	&	$-$0.006$\pm0.006$  &	8.67$\pm0.121$	    &	   &	       8.63	&    8.60 \\
1-24148  &  	    --- 	   & ---	   & 	& ---	  &	--- 	    &	   &	       8.79	&    8.80 \\
1-163966 &   $-$0.002$\pm0.001$    & 8.80$\pm0.008$& 	&$-$0.006$\pm0.001$   &	8.81$\pm0.012$ 	    &	   &	       8.71	&    8.71 \\
1-149561 &   ---		   &  ---	   & 	& ---	  &	--- 	    &	   &	       8.61	&    8.52 \\
1-295542 &    ---		   &  ---	   & 	& ---	  & ---	 	    &	   &	       8.55	&    8.45 \\
1-24660  &   $-$0.003$\pm0.002$    & 8.81$\pm0.033$ & 	&$-$0.007$\pm0.001$	  & 8.80$\pm0.018$  &	   &	       8.67	&    8.58 \\
1-258373 &   0.002$\pm0.001$    & 8.74$\pm0.011$ & 	& 0.001$\pm0.002$  & 8.78$\pm0.023$ &	   &	       8.61	&    8.54 \\
1-296733 &   ---		   & ---	   & 	& ---	  &	--- 	    &	   &	       8.65	&    8.66 \\
1-60653  &   0.0002$\pm0.001$    & 8.75$\pm0.014$ & 	& 0.0005$\pm0.001$	  &	8.73$\pm0.018$   &	   &	       8.56	&    8.52 \\
1-109056 &   $-$0.006$\pm0.003$    & 8.77$\pm0.036$ & 	&	$-$0.005$\pm0.001$ &8.66$\pm0.014$  &	   & 8.62	&    8.62 \\
1-210646 &   $-$0.006$\pm0.0004$    & 8.75$\pm0.004$ & 	& $-$0.010$\pm0.001$  &	 8.75$\pm0.011$ &	   &	       8.56	&    8.47 \\
1-248420 &   $-$0.005$\pm0.001$    & 8.82$\pm0.009$ & 	&$-$0.012$\pm0.003$	  &	 8.85$\pm0.025$ &	   &	       8.67	&    8.59 \\
1-277552 &   $-$0.007$\pm0.001$    & 8.75$\pm0.012$ & 	&$-$0.005$\pm0.001$	  &	8.68$\pm0.016$  &	   &	       8.58	&    8.44 \\
1-96075  &   $-$0.003$\pm0.001$    & 8.77$\pm0.006$ & 	&$-$0.007$\pm0.004$	  &	8.79$\pm0.034$  &	   &	       8.67	&    8.56 \\
1-558912 &   ----		   & ---	   & 	& ---	  & ---	 	    &	   &	       8.70	&    8.62 \\
1-269632 &   ---		   & ---	   & 	& ---	  &	--- 	    &	   &	       8.63	&    8.52 \\
1-258599 &   ---		   & ---	   & 	& ---	  &	--- 	    &	   &	       8.51	&    8.50 \\
1-121532 &   ---		   & ---	   & 	& ---	  &	--- 	    &	   &	       8.60	&    8.59 \\
1-209980 &   0.010$\pm0.005$	   & 8.46$\pm0.071$ & 	&0.003$\pm0.002$  &	8.48$\pm0.029$	    &	   &	       8.65	&    8.57 \\
1-44379  &   0.002$\pm0.002$    & 8.72$\pm0.025$ & 	&$-$0.002$\pm0.001$  &8.70$\pm0.011$    &	   &	       8.73	&    8.69 \\
1-149211 &   ---		   & ---	   & 	& ---	  & ---	 	    &	   &	       8.41	&    8.49 \\
1-279147 &   ---		   & ---	   & 	& ---	  &	--- 	    &	   &	       8.59	&    8.51 \\
1-94784  &   $-$0.002$\pm0.001$	   & 8.77$\pm0.010$ & 	&0.005$\pm0.001$  &	8.72$\pm0.011$    &	   &	       8.76	&    8.71 \\
1-339094 &   ---		   & ---	   & 	& ---	  &	--- 	    &	   &	       8.58	&    8.57 \\
1-137883 &   ---		   & ---	   & 	& ---	  & ---	 	    &	   &	       8.58	&    8.58 \\
1-48116  &   $-$0.0003$\pm0.001$	   & 8.81$\pm0.008$ & 	& 0.003$\pm0.001$	  &	8.76$\pm0.005$ &	   &	       8.61	&    8.54 \\
1-135641 &   $-$0.002$\pm0.001$    & 8.82$\pm0.017$ & 	&$-$0.001$\pm0.001$  & 8.82$\pm0.013$  &	   &	       8.66	&    8.78 \\
1-248389 &   ---		   & ---	   & 	& ---	  & ---	 	    &	   &	       8.82	&    8.89 \\
1-321739 &   $-$0.0004$\pm0.0003$    & 8.71$\pm0.013$ & 	&	$-$0.0004$\pm0.0004$  &	8.65$\pm0.013$    &	   &	       8.59	&    8.54 \\
1-234618 &   $-$0.0004$\pm0.0001$    & 8.74$\pm0.006$ & 	&$-$0.0004$\pm0.0001$  & 8.67$\pm0.008$  &	   &	       8.57	&    8.59 \\
1-351790 &   ---		   & ---	   & 	& ---	  &	--- 	    &	   &	       8.47	&    8.67 \\
1-23979  &   ---		   & ---	   & 	& ---	  &	--- 	    &	   &	       8.51	&    8.49 \\
1-542318 &   ---           & ---       &    & ---	  & ---         &	 &                 8.69	 &    8.62 \\
1-279676 &   $-$0.004$\pm0.003$    & 8.80$\pm0.031$ & 	&$-$0.008$\pm0.005$	  & 8.77$\pm0.054$ &	   &	       8.64	&    8.62 \\
1-519742 &   $-$0.017$\pm0.007$	   & 8.76$\pm0.039$ & 	& 0.010$\pm0.012$  & 8.56$\pm0.063$   &	   &	       8.59	&    8.63 \\
1-94604  &   $-$0.022$\pm0.008$    & 8.81$\pm0.063$ & 	&	$-$0.017$\pm0.003$  &	8.67$\pm0.020$ &	   &	       8.67	&    8.61 \\
1-37036  &   ---		   & ---	   & 	& ---	  & ---	 	    &	   &	       8.71	&    8.64 \\
1-167688 &   ---		   & ---	   & 	& ---	  &	--- 	    &	   &	       8.56	&    8.62 \\
1-279666 &   ---		   & ---	   & 	& ---	  &	--- 	    &	   &	       8.68	&    8.63 \\
1-148068 &   0.003$\pm0.001$    & 8.73$\pm0.017$ & 	& $-$0.006$\pm0.001$  &	8.83$\pm0.015$	    &	   &	       8.71	&    8.60 \\
1-603941 &   $-$0.008$\pm0.001$    & 8.77$\pm0.007$ & 	&$-$0.012$\pm0.001$	  &	8.79$\pm0.017$   &	   &	       8.61	&    8.54 \\
1-153627 &   $-$0.004$\pm0.0005$    & 8.80$\pm0.006$ & 	&$-$0.006$\pm0.001$	  &	8.75$\pm0.019$   &	   &	       8.55	&    8.58 \\
1-270129 &   0.003$\pm0.001$    & 8.74$\pm0.006$ & 	& 0.001$\pm0.002$  & 8.77$\pm0.020$	    &	   &	       8.67	&    8.56 \\
1-298938 &   $-$0.009$\pm0.001$    & 8.80$\pm0.017$ & 	&$-$0.008$\pm0.002$	  &	 8.70$\pm0.030$    &	   &	       8.61	&    8.54 \\
1-420924 &   $-$0.003$\pm0.001$    & 8.79$\pm0.014$ & 	& $-$0.002$\pm0.001$ &	8.78$\pm0.016$ &	   &	       8.63	&    8.55 \\
1-626658 &   $-$0.005$\pm0.002$    & 8.79$\pm0.015$ & 	& $-$006$\pm0.002$  & 8.79$\pm0.024$    &	   &	       8.67	&    8.60 \\
1-603039 &   $-$0.001$\pm0.001$	   & 8.81$\pm0.012$ & 	& 0.002$\pm0.0005$  &	8.74$\pm0.005$  &	   &	       8.60	&    8.51 \\
1-43868  &   0.017$\pm0.003$    & 8.64$\pm0.062$ & 	& $-$002$\pm0.001$  & 8.82$\pm0.021$	    &	   &	       8.70	&    8.74 \\
1-71987  &    ---		   & ---	   & 	& ---	  & ---	 	    &	   &	       8.62	&    8.57 \\
1-121973 &    ---		   & ---	   & 	& ---	  &	--- 	    &	   &	       8.75	&    8.68\\
1-122304 &    ---		   & ---	   & 	& ---	  &	--- 	    &	   &	       8.66	&    8.56\\ 
1-174631 &    ---		   & ---	   & 	& ---	  &	--- 	    &	   &	       8.49	&    8.58\\
\hline
\end{tabular}
\end{table*}

\begin{table*}
\contcaption{} 
\begin{tabular}{lcccccccc}
\hline
\noalign{\smallskip}

            &    \multicolumn{2}{c}{$N2$}               &    & \multicolumn{2}{c}{$O3N2$}   &      &      \multicolumn{2}{c}{12+log(O/H)}       \\
     ID     &     slope              & (O/H)$_{0}$      &    &  slope   & (O/H)$_{0}$	    &	   &	${\rm O/H})_{C}$  & $\rm (O/H)_{SB98f}$ \\
\cline{2-3} 
\cline{5-6}
\cline{8-9}
\noalign{\smallskip}
(1)         &     (2)               &     (3)           &   &  (4)      & (5)       	   &	   &		(6)	       &     (7)       \\	

1-617323    & $-$0.001$\pm0.002$ & 8.82$\pm0.020$  	&   & 0.003$\pm0.003$	&	8.77$\pm0.032$  &	   &	    8.64    & 8.62 \\	
1-176644    & $-$0.008$\pm0.004$    & 8.77$\pm0.053$ 	&   &$-$0.013$\pm0.003$	& 8.79$\pm0.044$ &	   &	    8.74  & 8.66  \\	
1-177972    & ---		    & ---	    	&   & ---	 	& ---	    	   &	   &	    8.55    & 8.55  \\	
1-179679    & 0.016$\pm0.004$     & 8.68$\pm0.029$ 	&   & $-$0.005$\pm0.002$ &	8.81$\pm0.014$ &	   &	    8.75    & 8.76  \\	
1-196597    & $-$0.0005$\pm0.002$      & 8.79$\pm0.040$ &   & $-$0.001$\pm0.001$& 8.72$\pm0.017$ &	   &	    8.66    & 8.61  \\	
1-210020    &$-$0.007$\pm0.003$     & 8.86$\pm0.075$ 	&   & $-$0.005$\pm0.001$& 8.70$\pm0.032$ &	   &	    8.59    & 8.56  \\	
1-201392    & ---		    & ---	    	&   & ---	 	& ---	    	   &	   &	    8.63    & 8.54  \\	
1-209707    & ---		    & ---	    	&   & ---	 	& ---	    	   &	   &	    8.51    & 8.51  \\	
1-209772    & ---		    & ---	    	&   & ---	 	& ---	    	   &	   &	    8.67    & 8.60  \\	
1-633942    & $-$0.006$\pm0.001$    & 8.82$\pm0.010$ 	&   & $-$0.005$\pm0.001$ &	8.77$\pm0.015$  &	   &	    8.74 & 8.70  \\	
1-277257    & 0.0005$\pm0.001$      & 8.72$\pm0.018$ 	&   & 0.004$\pm0.002$ 	& 8.59$\pm0.027$ &	   &	    8.60    & 8.63  \\	
1-298778    & 0.002$\pm0.001$      & 8.78$\pm0.004$ &   & 0.013$\pm0.005$ 	& 8.70$\pm0.032$  &	   &	    8.77    & 8.78  \\	
1-299013    & $-$0.012$\pm0.001$    & 8.85$\pm0.012$ 	&   & $-$0.018$\pm0.003$& 8.84$\pm0.026$ &	   &	    8.67    & 8.72  \\	
1-323794    & 0.001$\pm0.002$  & 8.66$\pm0.037$ 	&   & 0.001$\pm0.001$ & 8.59$\pm0.030$  &	   &	    8.54    & 8.53  \\	
1-384124    & ---		    & ---	    	&   & ---	 	& ---	    	   &	   &	    8.63    & 8.60  \\	
1-405760    & ---		    & ---	    	&   & ---	 	& ---	    	   &	   &	    8.61    & 8.58  \\	
1-625513    & $-$0.009$\pm0.001$    & 8.84$\pm0.016$ 	&   & 0.008$\pm0.001$ &	8.76$\pm0.023$ &	   &	    8.57    & 8.52  \\	
1-519412    & ---		    & ---	    	&   & ---	 	& ---	    	   &	   &	    8.45    & 8.46  \\	
1-547402    & $-$0.001$\pm0.003$      & 8.88$\pm 0.72$	&   & $-$003$\pm0.003$ 	& 8.59$\pm0.069$  &	   &	    8.47    & 8.52  \\	
1-175889    & 0.002$\pm0.001$      & 8.73$\pm0.010$ 	&   &  0.001$\pm0.002$	& 8.75$\pm0.030$ &	   &	    8.59    & 8.54  \\	
1-605353    & $-$0.008$\pm0.002$    & 8.84$\pm0.020$ 	&   & $-$0.011$\pm0.003$ & 8.82$\pm0.029$ &	   &	    8.77    & 8.86  \\	
1-232143    & $-$0.011$\pm0.001$    & 8.73$\pm0.013$ 	&   & $-$0.015$\pm0.001$ &	8.73$\pm0.010$ &	   &	    8.45    & 8.46  \\	
1-251458    & 0.001$\pm0.001$    & 8.77$\pm0.008$ 	&   & $-$0.001$\pm0.004$ & 8.82$\pm0.023$  &	   &	    8.69    & 8.64  \\	
1-298298    & $-$0.004$\pm0.001$    & 8.75$\pm0.011$ 	&   & $-$0.004$\pm0.001$ &	8.72$\pm0.013$  &	   &	    8.57    & 8.61  \\	
1-380097    & $-$0.002$\pm0.0004$    & 8.78$\pm0.003$ 	&   & 0.001$\pm0.003$ & 8.74$\pm0.023$  &	   &	    8.60    & 8.51  \\	
1-31788     & 0.002$\pm0.001$    & 8.79$\pm0.008$ 	&   & 0.0002$\pm0.002$ & 8.75$\pm0.022$ &	   &	    8.56    & 8.49  \\	
1-46056     & ---		    & ---	    	&   & ---	 	& ---	    	   &	   &	    8.54    & 8.54  \\	
1-114252    & ---		    & ---	    	&   & ---	 	& ---	    	   &	   &	    8.64    & 8.57  \\	
1-150947    & 0.008$\pm0.003$      & 8.67$\pm0.034$ 	&   & 0.010$\pm0.002$ &	8.57$\pm0.018$ &	   &	    8.55    & 8.50  \\	
1-604912    & 0.002$\pm0.004$      & 8.82$\pm0.056$ 	&   & $-$0.003$\pm0.003$ &	8.78$\pm0.037$ &	   &	    8.58    & 8.53  \\	
1-145679    & 0.003$\pm0.001$      & 8.74$\pm0.013$ 	&   & 0.003$\pm0.003$	& 8.76$\pm0.036$ &	   &	    8.78    & 8.69  \\	
1-163789    & $-$0.048$\pm0.002$    & 9.09$\pm0.017$ 	&   & $-$0.043$\pm0.008$ &	9.02$\pm0.071$ &	   &	    8.65    & 8.59  \\	
1-635348    & ---		    & ---	    	&   & ---	 	& ---	    	   &	   &	    8.76    & 9.00  \\	
1-153901    & ---		    & ---	    	&   & ---	 	& ---	    	   &	   &	    8.62    & 8.53  \\	
1-201969    & $-$0.014$\pm0.007$    & 8.88$\pm0.029$ 	&   & $-$0.006$\pm0.002$ & 8.95$\pm0.084$  &	   &	    8.73    & 8.66  \\	
1-196637    & ---		    & ---	    	&   & ---	 	& ---	    	   &	   &	    8.64    & 8.59  \\	
1-229862    & $-$0.001$\pm0.002$    & 8.80$\pm0.015$ 	&   & $-$0.012$\pm0.011$ &	8.83$\pm0.087$ &	   &	    8.66    & 8.65  \\	
1-229731    & 0.003$\pm0.0003$    & 8.74$\pm0.003$ 	&   & 0.0004$\pm0.001$	& 8.79$\pm0.009$  &	   &	    8.69    & 8.62  \\	
1-264729    & $-$0.001$\pm0.0004$    & 8.70$\pm0.023$ 	&   & $-$0.001$\pm0.0002$ &	8.63$\pm0.009$  &	   &	    8.54    & 8.53  \\	
1-268479    & $-$0.012$\pm0.003$    & 8.87$\pm0.030$ 	&   & $-$0.015$\pm0.002$ &	8.82$\pm0.021$ &	   &	    8.68   & 8.56  \\	
1-295041    & $-$0.002$\pm0.001$    & 8.80$\pm0.009$ 	&   & $-$0.020$\pm0.003$ & 8.84$\pm0.025$  &	   &	    8.58    & 8.54  \\	
1-281125    & ---		    & ---	    	&   & ---	 	& ---	    	   &	   &	    8.16    & 8.42  \\	
1-298498    & ---		    & ---	    	&   & ---	 	& ---	    	   &	   &	    8.31    & 8.41  \\	
1-297172    & ---		    & ---	    	&   & ---	 	& ---	    	   &	   &	    8.61    & 8.55  \\	
1-317962    & ---		    & ---	    	&   & ---	 	& ---	    	   &	   &	    8.58    & 8.47  \\	
1-318148    & ---		    & ---	    	&   & ---	 	& ---	    	   &	   &	    8.33    & 8.45  \\	
1-379811    & ---		    & ---	    	&   & ---	 	& ---	    	   &	   &	    8.51    & 8.47  \\	
1-605069    & ---		    & ---	    	&   & ---	 	& ---	    	   &	   &	    8.62    & 8.56  \\	
1-376346    & ---		    & ---	    	&   & ---	 	& ---	    	   &	   &	    8.63    & 8.55  \\	
1-382697    & ---		    & ---	    	&   & ---	 	& ---	    	   &	   &	    8.55    & 8.51  \\	
1-382452    & ---		    & ---	    	&   & ---	 	& ---	    	   &	   &	    8.59    & 8.56  \\	
1-403982    & $-$0.004$\pm0.006$    & 8.81$\pm0.050$ 	&   &  $-$0.001$\pm0.001$ &	8.70$\pm0.008$ &	   &	    8.68    & 8.68  \\	
1-605215    & $-$0.005$\pm0.001$    & 8.82$\pm0.007$ 	&   & $-$0.012$\pm0.001$ & 8.80$\pm0.011$ &	   &	    8.63    & 8.57  \\	
1-457424    & ---		    & ---	    	&   &	--- 	& ---	           &	   &	    8.72    & 8.79  \\	
1-537120    & ---		    & ---	   	    &   &	--- 	& ---              &	   &	    8.73    & 8.69  \\	

\hline
\end{tabular}
\end{table*}

\subsection{AGN calibrations}
\label{abund}

The O/H abundance in each AGN  was derived using its measured emission line
ratios and two calibrations, one proposed  by \citet{SB98} and the other recently proposed by \citet{carvalho19}. In what follows,  descriptions of these calibrations
are presented.

\subsubsection{\citet{SB98} calibration}

\citet{SB98}, by using a grid of photoionization models built with the {\sc Cloudy} code \citep{Ferland+17}, proposed two theoretical calibrations between  the emission line ratios
[\ion{N}{ii}]$\lambda$6584/H$\alpha$, [\ion{O}{iii}]$\lambda5007$/[\ion{O}{ii}]$\lambda3727$
and [\ion{O}{iii}]$\lambda5007$/H$\alpha$ and the metallicity (traced by the O/H abundance).  
These calibrations are valid for the range of $\rm 8.4 \: \leq \: 12+\log(O/H) \: \leq \:  9.4$ and the O/H abundances obtained from these calibrations differ by only $\sim0.1$ dex \citep{SB98, dors20b}.

In this work we used only one calibration proposed by \citet{SB98} ( hereafter SB$_{1}$) because  the 
[\ion{O}{ii}]$\lambda3727$
is not available in our data set. The SB$_{1}$ calibration is defined by:
 \begin{eqnarray}
       \begin{array}{l@{}l@{}l}
\rm 12+(O/H)_{SB_{1}} & = &  8.34  + (0.212 \, x) - (0.012 \,  x^{2}) - (0.002 \,  y)  \\  
         & + & (0.007 \, xy) - (0.002  \, x^{2}y) +(6.52 \times 10^{-4} \, y^{2}) \\  
         & + & (2.27 \times 10^{-4} \, xy^{2}) + (8.87 \times 10^{-5} \, x^{2}y^{2}),   \\
     \end{array}
\label{cal_SB}
\end{eqnarray}
\noindent where $x$ = [N\,{\sc ii}]$\lambda$$\lambda$6548,6584/H$\alpha$ and 
$y$ = [O\,{\sc iii}]$\lambda$$\lambda$4959,5007/H$\beta$.
 In order to take into account the dependence of this relation on the gas electron density ($N_{\rm e}$), we  applied the correction proposed by these authors:
\begin{equation}
\label{cal_SB2}
{\rm \log(O/H)_{SB98f}=[\log(O/H)_{SB_{1}}}]-\left[0.1 \: \times \: \log \frac{N_{\rm e}({\rm cm^{-3}})}{300 \: ({\rm cm^{-3}})}\right].
\end{equation}

The electron density ($N_{\rm e}$), for each object, was calculated from the 
[\ion{S}{ii}]$\lambda 6716/\lambda 6731$ line ratio, 
using the {\sc IRAF/temden} task and assuming an electron temperature of 10\,000 K.

\subsubsection{\citet{carvalho19} calibration}

\citet{carvalho19}, by using a diagram [\ion{O}{iii}]$\lambda5007$/[\ion{O}{ii}]$\lambda3727$ versus [\ion{N}{ii}]$\lambda$6584/H$\alpha$, compared observational data of 
 463 Seyfert~2 nuclei ($z\lesssim$ 0.4) with  photoionization model predictions
 built with the {\sc Cloudy} code \citep{Ferland+17}, which considered a wide range of nebular parameters.
 From this comparison, they obtained a semi-empirical calibration 
 between the $N2$=log([\ion{N}{ii}]$\lambda$6584/H$\alpha$) line ratio and the metallicity $Z$ given 
 by  
\begin{equation}
(Z/{\rm Z_{\odot}}) = 4.01^{N2} -0.07, 
\label{calib1_agn}
\end{equation}
valid for $0.3 \: \la \: (Z/{\rm Z_{\odot}}) \: \la \: 2.0$. 
The metallicity results obtained from the expression above can be converted into oxygen abundance by
\begin{equation}
12+\log({\rm O/H})_{C}=12+\log[(Z/{\rm Z_{\odot})} \: \times \: 10^{\log(\rm O/H)_{\odot}}],    
\end{equation}
where $\log(\rm O/H)_{\odot}=-3.31$ is the solar value \citep{alende01}.
The $N2$ index has an advantage  over other metallicity indicators
because  it involves emission lines with very close wavelength:  thus, $N2$
is not strongly affected by dust extinction and uncertainties produced
by flux calibration \citep{marino13} as the $O3N2$. 
However, due to the fact that $N2$ and $O3N2$ indices involve the ions
  $\rm N^{+}$, $\rm O^{2+}$ and $\rm H^{+}$ which have different ionization potentials,
 i.e. 29.60 eV, 54.93 eV and 13.6 eV, respectively, they have
 a   higher dependence  on the ionization degree of the gas phase than other indices, for
 instance, $N2O2$. Recently, \citet{kumari19} suggested that
$O3N2$ index can be used as a metallicity tracer of the diffuse ionized gas (DIG) and of  low-ionisation
emission regions [LI(N)ERs] in passive regions of galaxies  These authors also pointed out that the $O3N2$ could also be extended for metallicity estimates in Seyferts.
Further calibrations  involving the $O3N2$  
could produce additional gas phase metallicity estimations in Seyferts,   supporting the results obtained in the present work, hence it is used as \ion{H}{ii} region metallicity diagnostic.

 Regarding the electron density effect on the strong line calibrations, in general,  NLRs of AGNs present higher $N_{\rm e}$ values in comparison with the ones in the gas phase of \ion{H}{ii} regions. For instance, \citet{copetti00}, adopting the
[\ion{S}{ii}]$\lambda6716/\lambda6731$ as a $N_{\rm e}$ indicator,
derived electron density values for a sample of galactic \ion{H}{ii} regions in the range of $N_{\rm e}\approx20-400$ cm$^{-3}$ (see also, \citealt{krabbe14, mora19, buzzo21, rosa21}). On the other hand, electron density estimates in AGNs based on
[\ion{S}{ii}]$\lambda6716/\lambda6731$ and [\ion{Ar}{iv}]$\lambda4711/\lambda4740$
line ratios by \citet{congiu17} show values ranging from $\sim200$ to 13\,000 cm$^{-3}$ 
(see also, \citealt{izabel18, revalski18, kakkad18, mingozzi19, Davies20}).
 It is worthwhile to state that, some emission lines with low  critical density $N_{\rm c}$
(e.g., the [\ion{O}{ii}]$\lambda$3727 and [\ion{S}{ii}]$\lambda6716$
lines have $N_{\rm c}=10^{3.7}$ cm$^{-3}$ and 1600 cm$^{-3}$ respectively; \citealt{vaona12}) are used in AGN strong line methods 
\citep{SB98, castro17} and they can suffer collisional de-excitation. Therefore, it is more important to consider  electron density
effects on  oxygen abundance obtained through  strong-line
methods for AGNs rather than  for \ion{H}{ii} regions.\\
In any case, the empirical \ion{H}{ii} region calibrations proposed
by \citet{Perez+09} and the semi-empirical AGN calibration proposed
by \citet{carvalho19} are obtained by using  observational data of 
large samples of objects, with a wide range of nebular parameters.
Therefore, in these calibrations, effects of electron density $N_{\rm e}$ (as well as reddening correction,
gas ionization degree, etc) on emission-line ratio intensities are considered in intrinsic way.
On the other hand, the theoretical \citet{SB98} calibration was based on photoionization
models, which the electron density was an input parameter. Therefore,  
these authors proposed a correction technique to take into account the $N_{\rm e}$ effect on the resulting O/H abundances. However, since typical $N_{\rm e}$ values in NLRs 
are around 500 $\rm cm^{-3}$ (see, for instance, Fig.~2 of \citealt{{dors21ep}}) the use of Eq.~\ref{cal_SB2} implies a correction in the total oxygen
abundance  of only $\sim$0.02 dex. Moreover, even considering the highest $N_{\rm e}$ value
found by \citet{dors21ep} for a large sample of Seyfert nuclei, i.e. $N_{\rm e}=2250 \: \rm cm^{-3}$, the O/H correction according to Eq.~\ref{cal_SB2}
is $\sim0.1$ dex, i.e. lower than the uncertainty of $\sim0.2$ dex in abundance estimates derived from strong-line methods (e.g. \citealt{SB98, DTT02, marino13}).
Thus, electron density has a marginal effect on the abundances derived from our sample.

We define  the discrepancy  $D$ (in dex) as being the difference 
between the O/H abundance of the AGN (derived from  \citealt{SB98} or \citealt{carvalho19} calibrations)  and the intersect oxygen abundances  derived from the radial abundance gradients:
\begin{equation}
\label{newdf}
D=\rm [12+\log(O/H)_{AGN}]-[12+log(O/H)_{0}],    
\end{equation}
where ${\rm 12+\log(O/H)_{0}}\equiv Y_{0}$.

\section{Results and Discussion}
\label{resdic}

\begin{figure}
\includegraphics[width=\columnwidth]{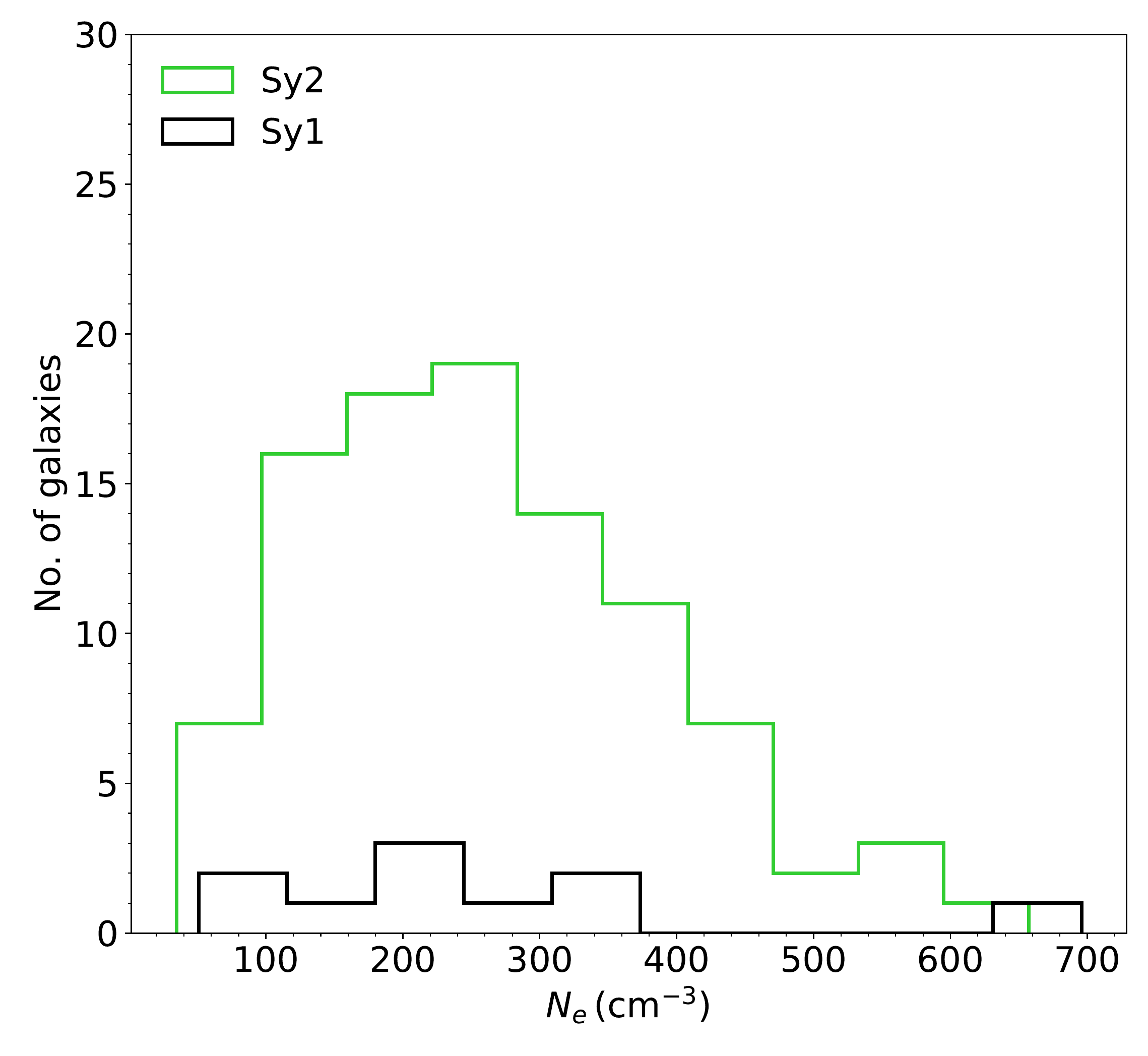}
\vspace*{-2mm}
\caption{Electronic density distributions obtained within the 2.5 arcsec central region for the 107 AGN of our sample. The distribution for 
Seyfert~1 and Seyfert~2 are represented in black and green
colors, respectively.}
\label{hist}
\end{figure}

In Fig.~\ref{hist} we present the electron density distribution obtained within the 2.5 arcsec central region, for the 108 Seyfert galaxies. The black 
and green distributions represent the results for Seyfert~1 and Seyfert~2 nuclei, respectively. The maximum and minimum values, taking into account the total sample of Seyfert~1 and Seyfert~2, are 696 cm$^{-3}$ and 35 cm$^{-3}$ respectively, with the
 mean value of $N_{\rm e}$ $\approx$ 260 cm$^{-3}$ and a median value of $N_{\rm e}$ $\approx$ 240 cm$^{-3}$. The
range of $N_{\rm e}$ values derived from our sample is
 somewhat lower than the $N_{\rm e}$ values obtained by \citet{Dors+20},
 who derived the range $100 \: \la \: N_{\rm e} \: (\rm cm^{-3}) \: \la \: 10\,000$  and an average value of $\sim 650$ cm$^{-3}$,  for a sample of 463 Seyfert~2 nuclei whose data were taken from the SDSS DR7 \citep{york00}. However,  only 15\,\% of the objects considered by \citet{Dors+20} present
 higher $N_{\rm e}$ values than the maximum value derived from our sample. \citet{vaona12} selected   about 
 2\,700 spectra of Seyfert nuclei from SDSS DR7 \citep{york00} and,
 among other nebular properties, estimated $N_{\rm e}$
 mainly based on the [\ion{S}{ii}]$\lambda6716/6731$ line ratio. These authors derived a range of values similar to the estimations  by \citet{Dors+20} and an average value
 of $N_{\rm e} \sim 250$ cm$^{-3}$, whereby most of the objects ($\sim2\,300$ galaxies) show values lower than 500 cm$^{-3}$, while higher electron density values greater than 1000 cm$^{-3}$  were derived  from  only  97 ($\sim 3\,\%$) objects  
 (see also \citealt{zhang13}). Thus, the discrepancy in $N_{\rm e}$ values above
   is  probably due to  the distinct and larger sample of objects considered by \citet{Dors+20} and \citet{vaona12}.
  It is worth to mention that the  $N_{\rm e}$ values obtained from our sample  are well below the critical density value
($N_{\rm c}$) of the emission-lines considered in the present work. In fact, the emission line with the 
lowest  $N_{\rm c}$ is   [\ion{S}{II}]$\lambda$6716
($N_{\rm c} \sim 1600 \rm \: cm^{-3}$,  \citealt{vaona12}), indicating that the collisional de-excitation is negligible in this case, which also does not affect the emissivity of the lines and, consequently the chemical abundance estimations (for a detailed discussion on electron density effects on
AGN abundance determinations see \citealt{dors20a}).

\begin{figure*}
\includegraphics[width=\linewidth]{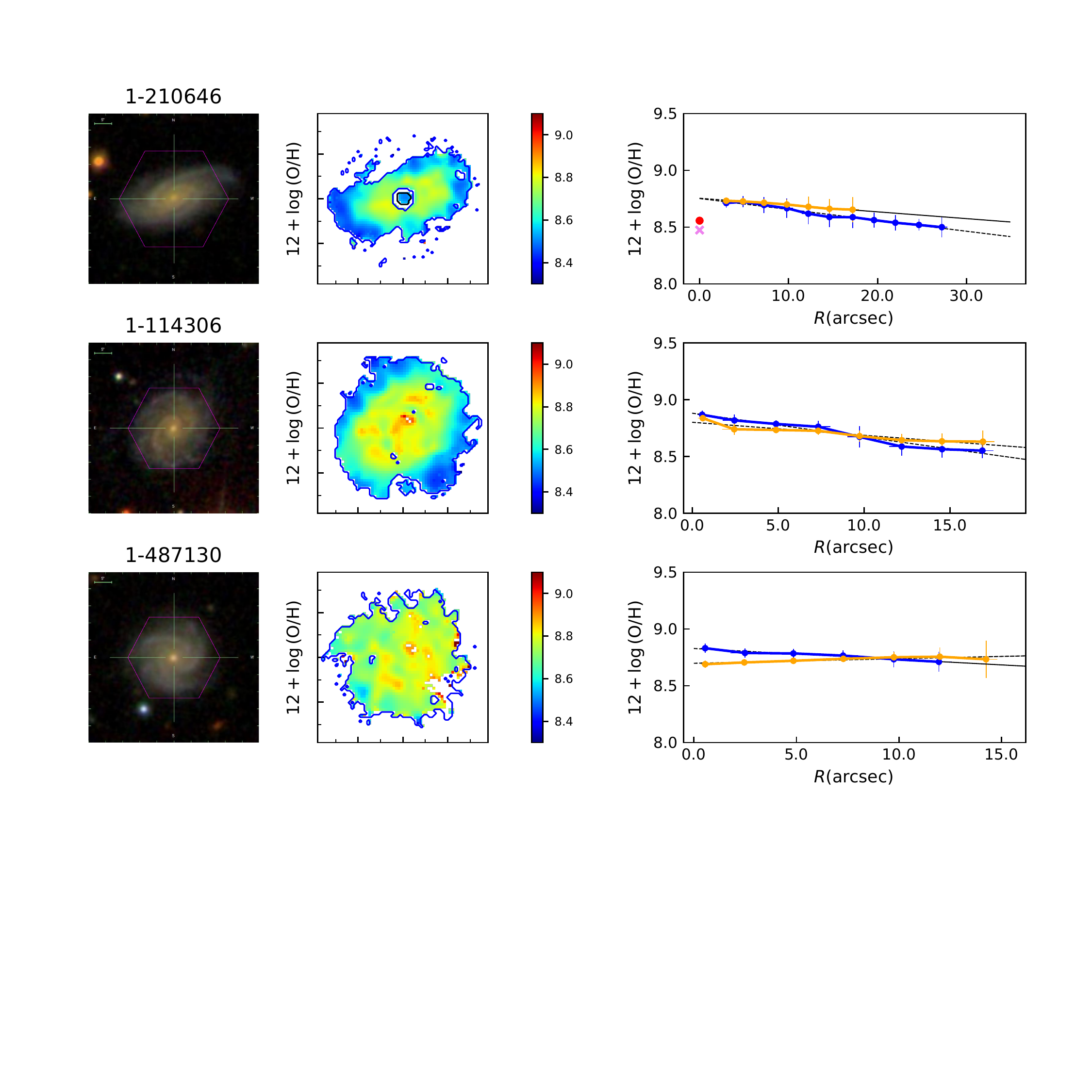}
\vspace*{-48mm}
\caption{Surface distribution of the gas abundances for the Seyfert~2 galaxy MANGA ID 1-210646 (top row) and its two controls (bottom rows). The  tick marks in the maps are separated by 5$^{\prime\prime}$. 
The left column shows the SDSS-IV images in the  bands \textit{gri} with the MaNGA footprint over-plotted in pink for the AGN (top) and its controls (bottom); central column: the metallicity distribution derived assuming the \citet{carvalho19} calibration for the AGN region (top row) and the \citet{Perez+09} calibrations for the HII regions in the disk of both the AGN and control galaxies (bottom rows); right column: mean azymuthal profile and standard deviations of the oxygen abundance [in units of 12+log(O/H)]. The profile in blue was obtained from Eq.~\ref{abund_HII} and the one in orange was obtained from Eq.~\ref{abund_HII_PMC09}, and are plotted
versus the galactocentric distance $R$ (in units of arcsec). In the top panel showing the AGN gradient, the central values obtained from the calibration proposed by \citet{carvalho19} and \citet{SB98} are shown  in red point and pink  cross sign point, respectively.}
\label{fig_2}
\end{figure*}

We  derived spatially resolved maps and radial chemical abundance profiles for the AGN together with controls and showing the corresponding maps for the galaxy with MaNGA ID 1-210646 (AGN) and its controls in Fig.~\ref{fig_2}. The maps for the remaining objects are shown in  Figures A1 - A72 in the Appendix. The radial profiles were corrected for projection, using information about the major and minor axis (from SDSS galaxy image, obtained from the MaNGA's drpall table) to determine the inclination of the galaxies. The inclination angle of the AGN sample and their controls range from 1 to 70 degrees (the control galaxies were chosen to match each of the selected AGN hosts). We were able to obtain the O/H radial gradients for 61 galaxies containing AGNs (from a total sample of 108 galaxies hosting AGNs)  and for 112 control galaxies (from a sample of 145 control galaxies). The chemical abundance profiles shown in blue and orange in Figs.~\ref{fig_2} and A1 - A72 were derived by using the calibrations (Eqs.~\ref{abund_HII} and ~\ref{abund_HII_PMC09}) from  \citet{Perez+09}. The dashed lines represent the fits of the points by   Eq.~\ref{graf}. The blue and orange points represent the mean values, with standard deviations, considering all spaxels (both from control and AGN galaxies) divided in bins of 2.5 arcsec 
width. The slopes of the gradients in each galaxy  and  extrapolated abundances [$Y_{0} \rm \equiv 12+log(O/H)_{0}$] to the center (galactocentric distance $R=0$) are listed in Table~\ref{table1}. 

\begin{figure*}
\centering
\includegraphics[width=0.95\textwidth,height=0.95\textheight, keepaspectratio]{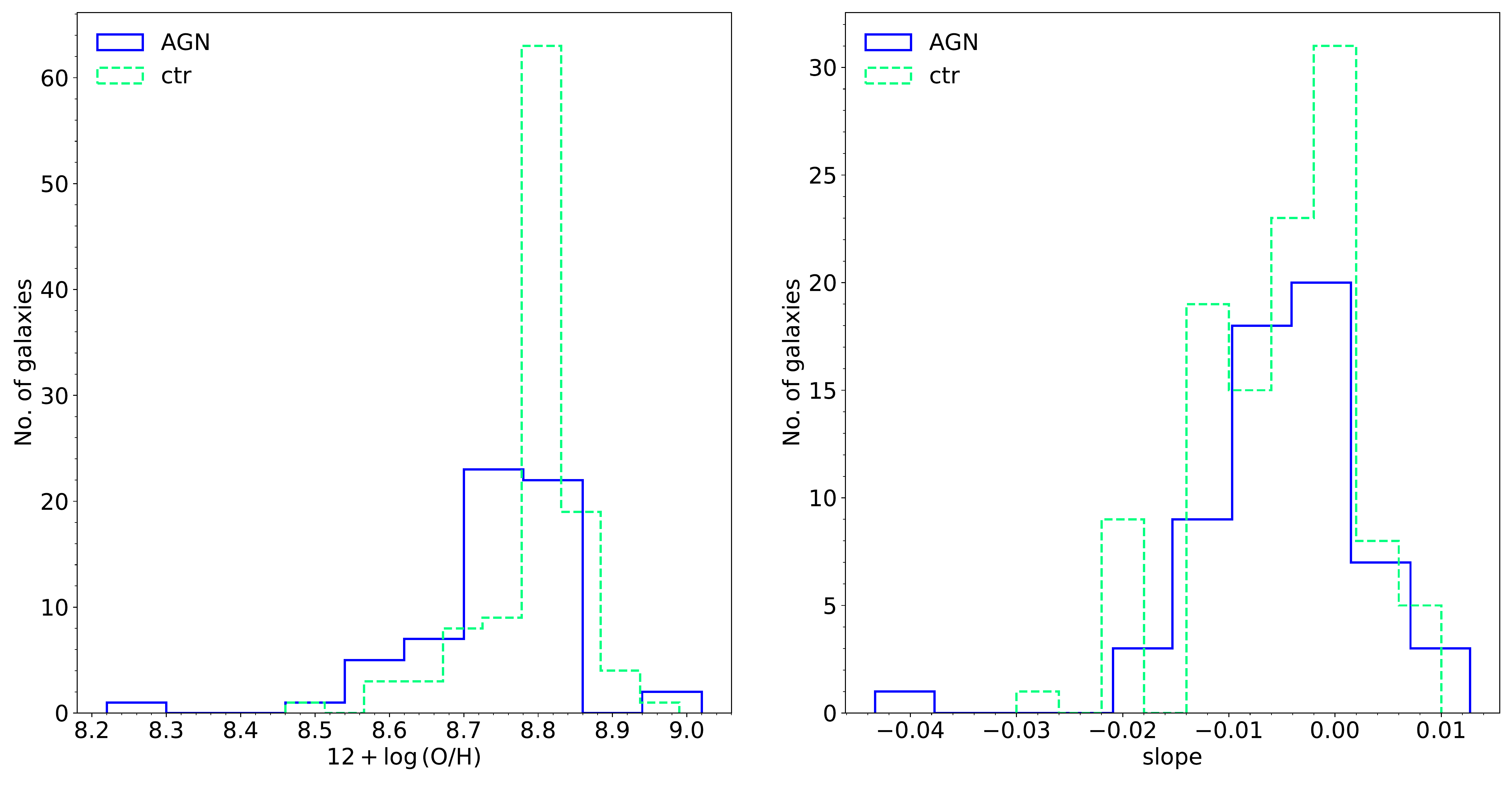}
\vspace*{-2mm}
\caption{Left: distribution of nuclear chemical abundance obtained using the extrapolation method for AGNs and their control galaxies. Both were obtained from the use of the $N2$ calibration in the \ion{H}{ii} regions. Right: histogram of the slopes of the abundance gradients for the AGNs compared to those from their control galaxies. AGN are shown in blue and the controls in light green dashed lines.}
\label{hist3}
\end{figure*}


Fig.~\ref{fig_2} and Figs. A1 - A72 show, in the left panels, {SDSS  image in the bands \textit{gri}   of each galaxy 
 with the MaNGA footprint over-plotted in pink}, and in the central panels, maps of the oxygen abundance distribution [in units of 12+log(O/H)] along the galactic disks.

The white regions in these maps represent the spaxels classified as transition objects or LI(N)ER-like objects (see also \citealt{kaplan16}) for which the calibrations used in this work  can not be applied. Although recent studies  \citep{kumari19, wu20, wu21} have investigated metallicity derivation in these object class, we did not obtain any estimate for them.
For  about 43\,\% of the AGN sample it was not possible
to obtain the radial abundance gradient  due to the fact that the spaxels in the disk  are not classified as SF-like objects in the diagnostic diagram (see Fig.~\ref{fig:bpt}), i.e. they are
classified as transition or AGN-like objects. These objects may include disk regions with 
a composite ionization source, such as shocks mixed with ionization from a young stellar cluster
(e.g. see  \citealt{allen08, rosa14}).
 Transition objects are indeed expected to have a stellar cluster--AGN mixing as their ionizing source (e.g. \citealt{kewley06, wu07, davies14}). Although recent studies have proposed methods to estimate the metallicity  in this class of objects (i.e. composite objects,  see for instance \citealt{wu20, wu21}), 
we did not estimate the abundances for them  because the main goal
of this work is the comparison between AGN and \ion{H}{ii} region metallicities.

Regarding the metallicity gradients presented in the right panels of Fig~\ref{fig_2} and Figs. A1 - A72 (for the AGN host disk and its control galaxies), we can state that: in the AGN host disk, the metallicity gradient is comparable to that found in the control galaxies; although we have used two different calibrations for the \ion{H}{ii} regions -- the blue profile derived from the $O3N2$ index (Eq.~\ref{eq_O3N2}) and the orange profile from the $N2$ index (Eq.~\ref{abund_HII_PMC09}) -- show similar gradients that are mostly negative. Our main findings are:
\begin{itemize}
    \item $\sim$66\,\% of the galaxies containing
AGNs  have negative radial gradients when these are derived via 
$O3N2$ and $N2$ indices (i.e.  the O/H abundance decreases as the galactocentric distance increases);
    \item $\sim$25\,\% of AGN hosts have positive gradients, indicating an increase in metallicity as we move away from the central region; and
    \item $\sim$9\,\%  of the AGN host sample present flat gradients.
\end{itemize}
 These results are confirmed by the histograms shown in Fig.~\ref{hist3}, comparing the central extrapolated abundances and the gradient slopes of the AGN hosts (blue line) 
 with those from their control galaxies (light green dashed lines).

\begin{figure*}
\includegraphics[width=0.95\textwidth,height=0.95\textheight, keepaspectratio]{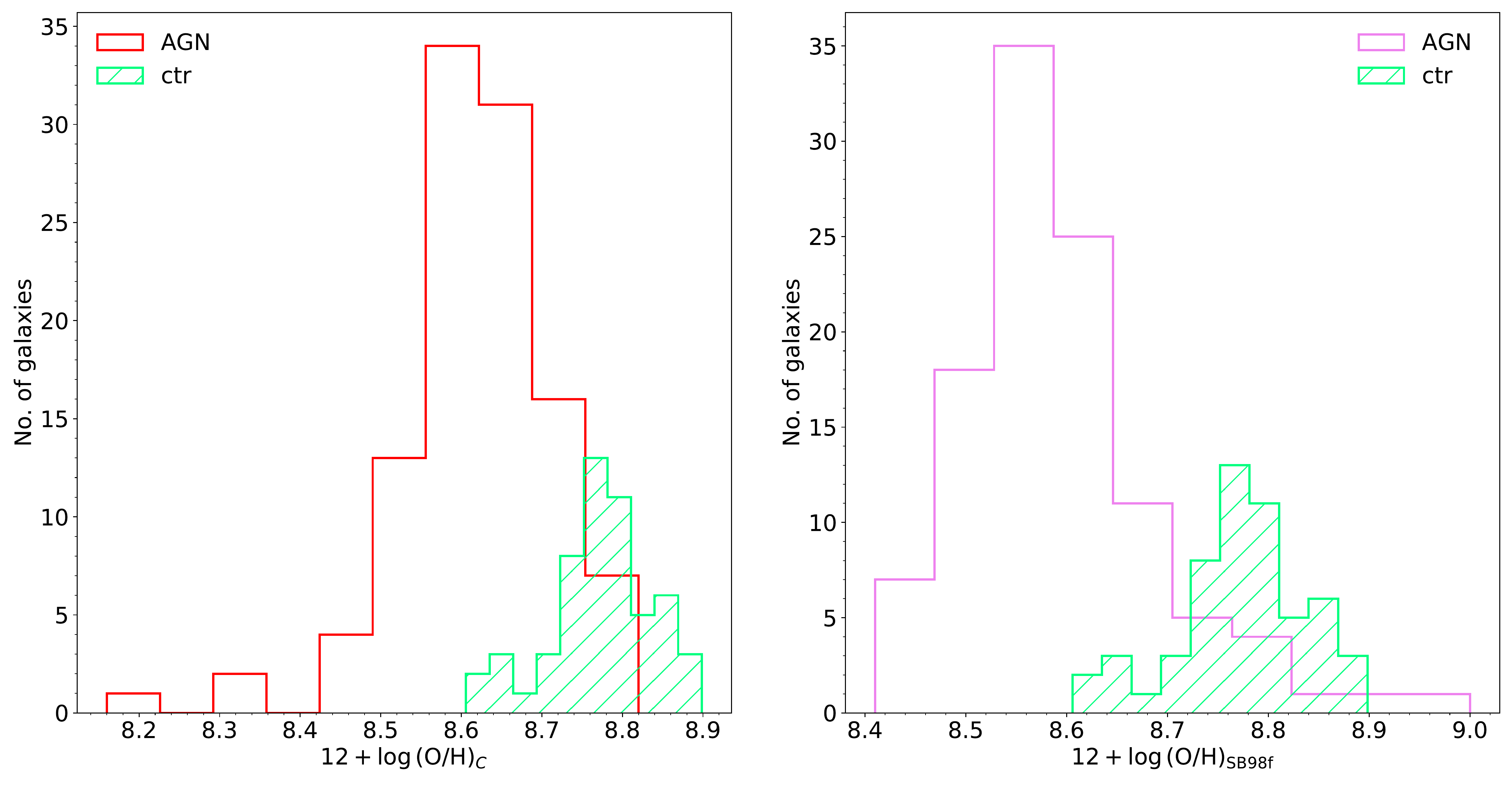}
\vspace*{-2mm}
\caption{Histograms showing the central oxygen abundances distribution for the AGNs compared to that of the control galaxies. In the left panel, the AGN abundances were obtained using the \citet {carvalho19} calibration (red open histogram) and in the right panel, using the \citet{SB98} calibration (pink open histogram). The control galaxies are represented by the hatched histogram in light green.}
\label{hist2}
\end{figure*}

If galaxies evolve as a closed system and are formed according to the
inside-out scenario (e.g. \citealt{molla05}), negative abundance gradients are expected. In fact, this feature is commonly found in most 
disks of spiral galaxies (e.g. \citealt{shaver83, vanzee98, kennicutt03, leonid04, dors05, sanchez14, belfiore17, evan20, mingozzi20}).
However, some relevant processes could affect the galaxy evolution resulting
in radial gradients which differ from the expected negative abundance gradients. For instance, some  galaxies are formed and evolve in a complex environment (galaxies can have a nurture evolution;
see \citealt{paulino19} and references therein), with gas circulating inside, outside and around of them, making galaxies evolve 
in short time scales \citep{Somerville+15}. This cycle includes  modified star-formation, accretion, mergers, and the way each
action impacts the galaxy is unique. If metal-poor gas accretion is deposited directly into the centers of galaxies, it should act to dilute the central metallicity, flatten or push down negative gradients  \citep{Simons+20}, producing a break in the gradients at small galactocentric distances.

 The histograms in Fig.~\ref{hist2} show the oxygen abundance distributions calculated within the inner 2.5 arcsec of each galaxy, i.e for the AGNs and their control galaxies (which have star-forming nuclei), whose O/H values were derived by using the calibrations described in 
Sect.~\ref{meth}. In the left and right panels
of Fig.~\ref{hist2} the AGNs abundances calculated by Eqs.~\ref{cal_SB2} and \ref{calib1_agn} 
are represented in red and pink colours, respectively.
 Distributions for the control sample are represented by the hatched area (in light green) and the abundances are estimated using Eq.~\ref{eq_O3N2}. 
In both panels of Fig.~\ref{hist2}, it can be seen that the 
results indicate that control galaxies present O/H abundance values higher than those derived for AGNs.

\begin{figure*}
\centering
\includegraphics[scale = 0.4]{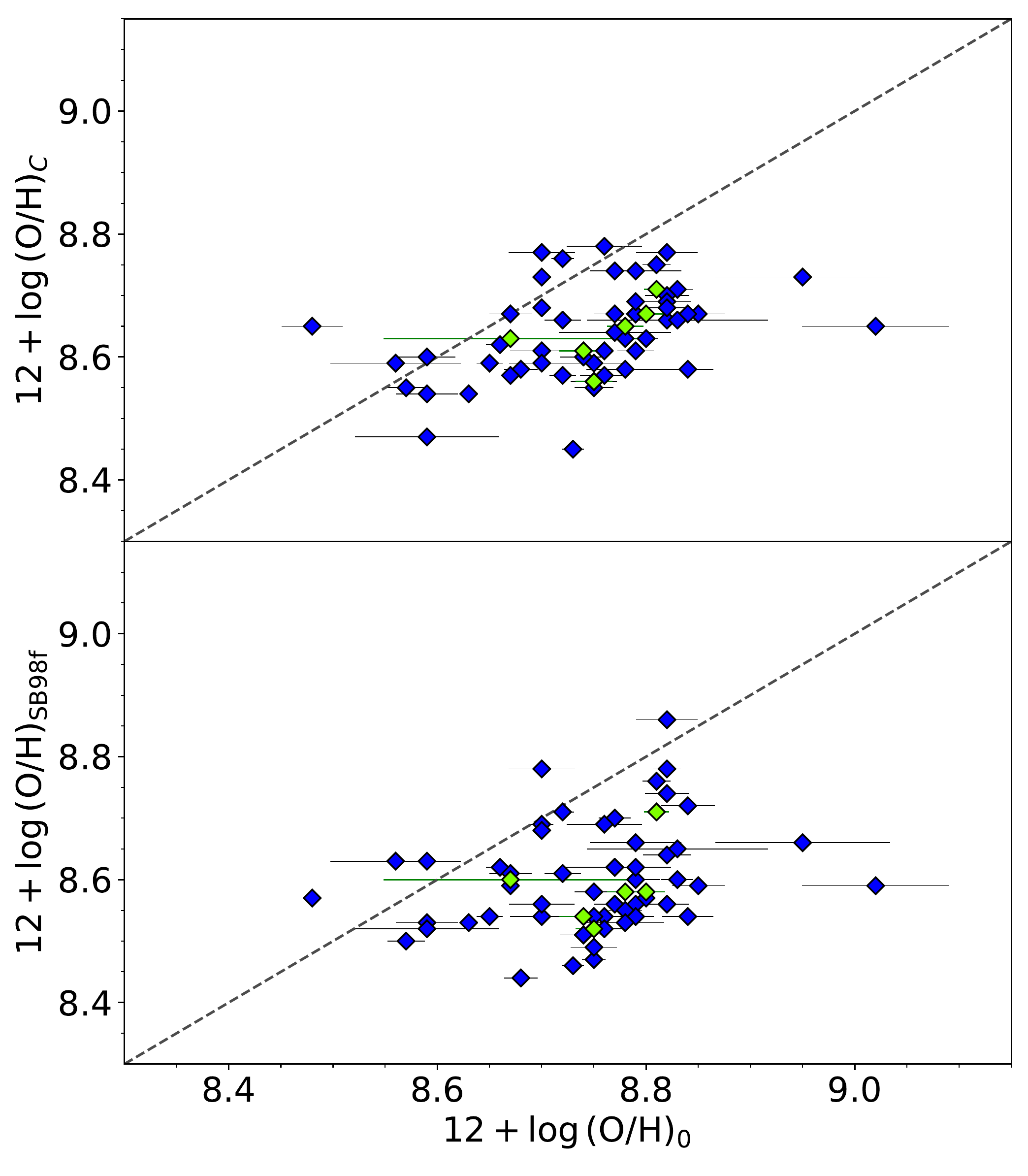}
\includegraphics[scale = 0.4]{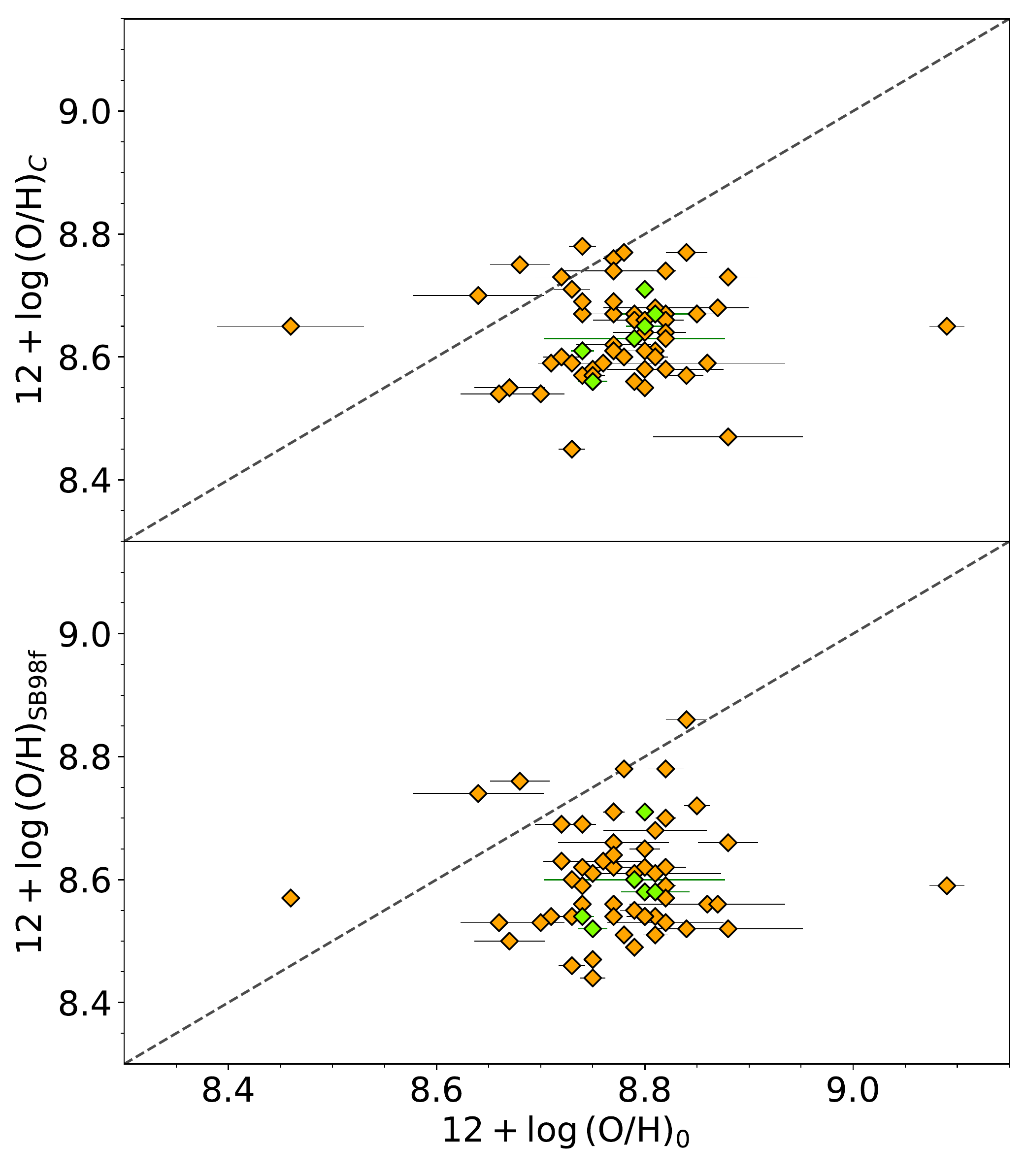}
\vspace*{-2mm}
\caption{Left panels: $\rm 12 + \log(O/H)$ calculated from \citet{carvalho19} (top panel) and  \citet{SB98} (bottom panel) vs. $\rm 12 + \log(O/H)_{0}$ calculated from the O3N2 calibration for the \ion{H}{ii} regions of PMC09. Right panels: $\rm 12 + \log(O/H)$ calculated from  \citet{carvalho19} (top panel) and  \citet{SB98} (bottom panel) vs.
$\rm 12 + \log(O/H)_{0}$ calculated from the N2 calibration for the \ion{H}{ii} regions of PMC09. The solid line represents the equality between the estimates.}
\label{graf_1}
\end{figure*}

In Fig.~\ref{graf_1}, the $\rm 12 + \log(O/H)$ values obtained from the AGN calibrations (see Sect.~\ref{abund}) are plotted against those inferred from the gradient extrapolations. The  $Y_{0}$ values (see Eq.~\ref{graf})
of the metallicity gradients were determined from the \ion{H}{ii} region calibrations based on $O3N2$ (left panels) and $N2$ (right panels) indices proposed by
\citet{Perez+09}. In general, it is possible to verify that, even considering the errors in the estimates, i.e. in order of $\pm0.1$ dex
(e.g. \citealt{DTT02, marino13}), the direct estimations for AGNs  (derived through  \citet{SB98} and \citet{carvalho19} calibrations) are lower than the estimates obtained from the gradient extrapolations for the high metallicity
regime ($\rm 12+\log(O/H) \: \ga \: 8.6$).
Our findings reveal that:
\begin{itemize}
    \item $\sim$80\,\% of the metallicity values of AGNs,
    derived from the \citet{SB98} and \citet{carvalho19} calibrations  are lower than those obtained
    via extrapolation methods (using both the $N2$ and $O3N2$ relations for the \ion{H}{ii} regions to derive the disk gradients);
    \item $\sim$15\,\% of the metallicity values 
   obtained for the AGNs are similar (within the uncertainties) to those obtained via the extrapolation estimates; and
    \item for $\sim$5\,\% of the sample we found the AGN metallicities derived from the calibrations above higher than the values derived from the extrapolations. 
    
\end{itemize}

\begin{figure*}
\includegraphics[scale=0.38]{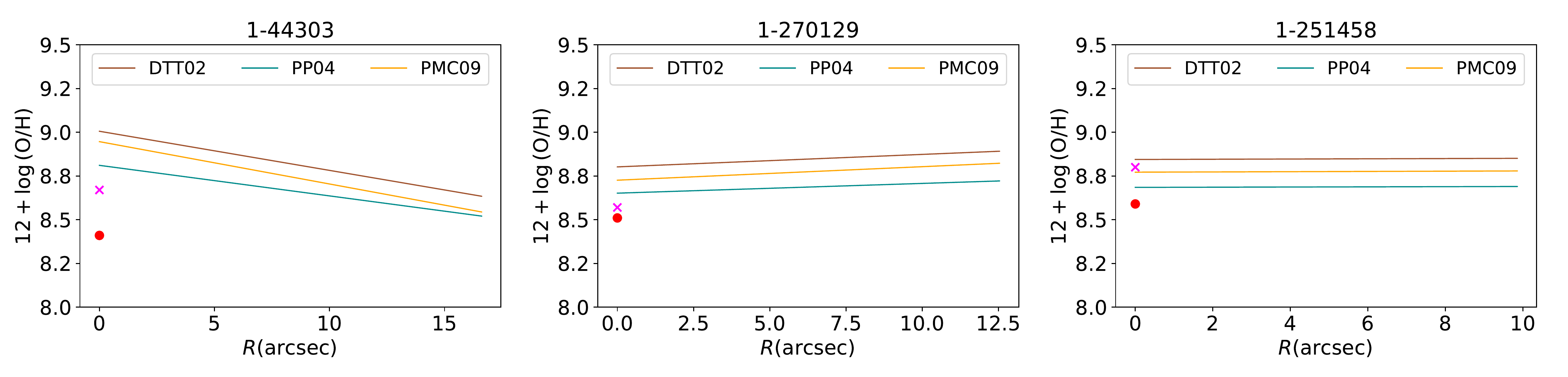}
\vspace*{-2mm}
\caption{Radial profiles of $\rm 12 + \log(O/H)$ obtained by using three different metallicity calibrations for star-forming regions. The lines represent linear
regression to the points (not shown):
the brown, dark cyan and orange lines represent linear regressions to the O/H abundance estimates for disk \ion{H}{ii} regions of each object obtained assuming the calibration by \citet{DTT02} - DTT02 (Eq.~\ref{abund_HII_DTT02}),  \citet{PP04} - PP04 (Eq.~\ref{abund_HII_pp04}) and   \citet{Perez+09} - PMC09 (Eq.~\ref{abund_HII_PMC09}), respectively. The red point and pink cross sign  points represent the abundances for AGN calculated via  \citet{carvalho19} (Eq.~\ref{calib1_agn})
and \citet{SB98} (Eq.~\ref{cal_SB}) calibrations, respectively.
The error in O/H abundances derived from strong emission line
line calibrations is of the order of $\pm0.1$ dex \citep{DTT02}.}
\label{fig_3}
\end{figure*}

 In Fig.~\ref{fig_2}, both the O/H radial gradients and the estimates for the AGN
(based on \citealt{SB98} and \citealt{carvalho19} calibrations)
for the Seyfert~2 galaxy  with MANGA ID 1-210646 are shown.
  It can be seen that a break 
in the radial gradient at small galactocentric distance is observed when considering the values based on the AGN calibrations,   while the difference between the two values (or discrepancy) $D$   is of the order of 0.5 dex  (see also Fig.~\ref{fig_3}).
In summary, there is a clear tendency of the AGN 
oxygen abundance (or metallicity) to be lower than the extrapolated O/H value for the high metallicity regime.

The above result is in agreement with some results found in previous studies. For instance,
\citet{Dors+15}, who  used long-slit spectroscopic data obtained by \citet{ho97},
found that the metallicity in the NLR  (derived from \citealt{SB98} calibration)
and in star-forming nuclei, whose  metallicity was estimated by using
the $C$-method \citep{leonid12, leonid13},  are  close to or slightly lower  than those obtained by the extrapolation method in the regime of high metallicity, i.e. 12+log(O/H)$\ga\: 8.6$.  These authors pointed out that metal-poor gas accretion is less evident for galaxies with
low-metallicity, where the metallicity of the accretion material is
similar (or not so different) to that of the gas phase in the central regions,
therefore, not producing significantly metallicity change. The opposite, probably, occurs for AGNs with high metallicity, where the infalling poor metallicity gas and the one in the galaxy differ considerably.
This scenario explains the result shown in Fig.~\ref{graf_1}.
\citet{sanchez14} found, for $\sim$10\,\% of the objects of the their sample, observational evidence of lower central oxygen abundances than those inferred from the O/H gradient extrapolation to the central parts of spiral galaxies  and suggested that a possible explanation is the addition of metal-poor gas to the center of the galaxies.

To verify if the result   shown in Fig.~\ref{graf_1} is dependent on the 
calibration assumed to derive the O/H gradients in our sample,
other empirical calibrations   were also considered.  In particular, we considered the empirical calibrations proposed by \citet{DTT02}:
\begin{equation}
\mathrm {12 + \log(O/H)} = 9.12 + 0.73 \times N2,
\label{abund_HII_DTT02}
\end{equation}
 and by \citet{PP04}:
\begin{equation}
\mathrm {12 + \log(O/H)} = 8.90 + 0.57 \times N2, 
\label{abund_HII_pp04}
\end{equation}
and calculated the radial gradients for all objects   in our sample.

In Fig.~\ref{fig_3}, we show the abundance gradients as well as the nuclear abundance values for three AGN hosts, with MANGA ID's 1-44303, 1-270129 and 1-251458, which are representative of the abundance profiles observed in our sample.
The resulting O/H gradients based on the three calibrations above
are represented by lines with different colours. These gradients were extrapolated to the
zero galactocentric distance ($R=0$). In  Fig.~\ref{fig_3}, the 12+log(O/H) values obtained for the AGNs using the \citet{carvalho19} and  \citet{SB98} calibrations are also indicated as a red dot and purple cross, respectively.
It can be seen that, independently  from the calibration considered
to derive the gradients in the AGN,  with exception of the object
1-251458, for which the \citet{SB98} calibration produces a high O/H value, we confirm that the extrapolated values are usually higher than those derived from the AGN calibrations.
Moreover, the error associated  with O/H estimates from calibrations based on strong
emission lines is in the order of $\pm$0.1 dex
(e.g. \citealt{DTT02, marino13}), lower than the average discrepancy $<D>$ (see below) derived for our sample. The rest of the objects (not shown) were subjected to the same process as those in Fig.~\ref{fig_3} and yielded similar results.


As discussed previously, there are several physical processes that can produce 
a decrease (or different O/H abundances)  in the O/H abundance (or $Z$) in the central parts of galaxies in comparison with that obtained from the gradient extrapolation. The simplest scenario appears to be the accretion of metal-poor gas into the nuclear region, e.g. via the capture of a gas rich dwarf galaxy, which is a process that can lead to the triggering of nuclear activity in galaxies, as discussed in \citet{thaisa_allan19}. A molecular (e.g. \citealt{riffel08, riffel13},  \citealt{moire18}) and/or neutral gas (e.g. \citealt{allison15}) reservoir can thus form in the surroundings of the supermassive black hole (SMBH), feeding it. 

Some recent studies have pointed out that star formation inside AGN gas outflows (e.g. \citealt{gallagher19}) and/or supernova explosions in accretion disks  can  locally enrich the AGN and produce very high abundances, mainly in the Broad Line Regions (e.g. \citealt{wang11, moranchel21}). But, in our case, we find low metallicity in the NLRs of the sample, and therefore, the most likely process is that we are observing an accretion of  poor metal gas rather than a higher star formation rate  in the innermost disk \ion{H}{ii} regions. 
 
\begin{figure*}

\includegraphics[width=1.02\columnwidth]{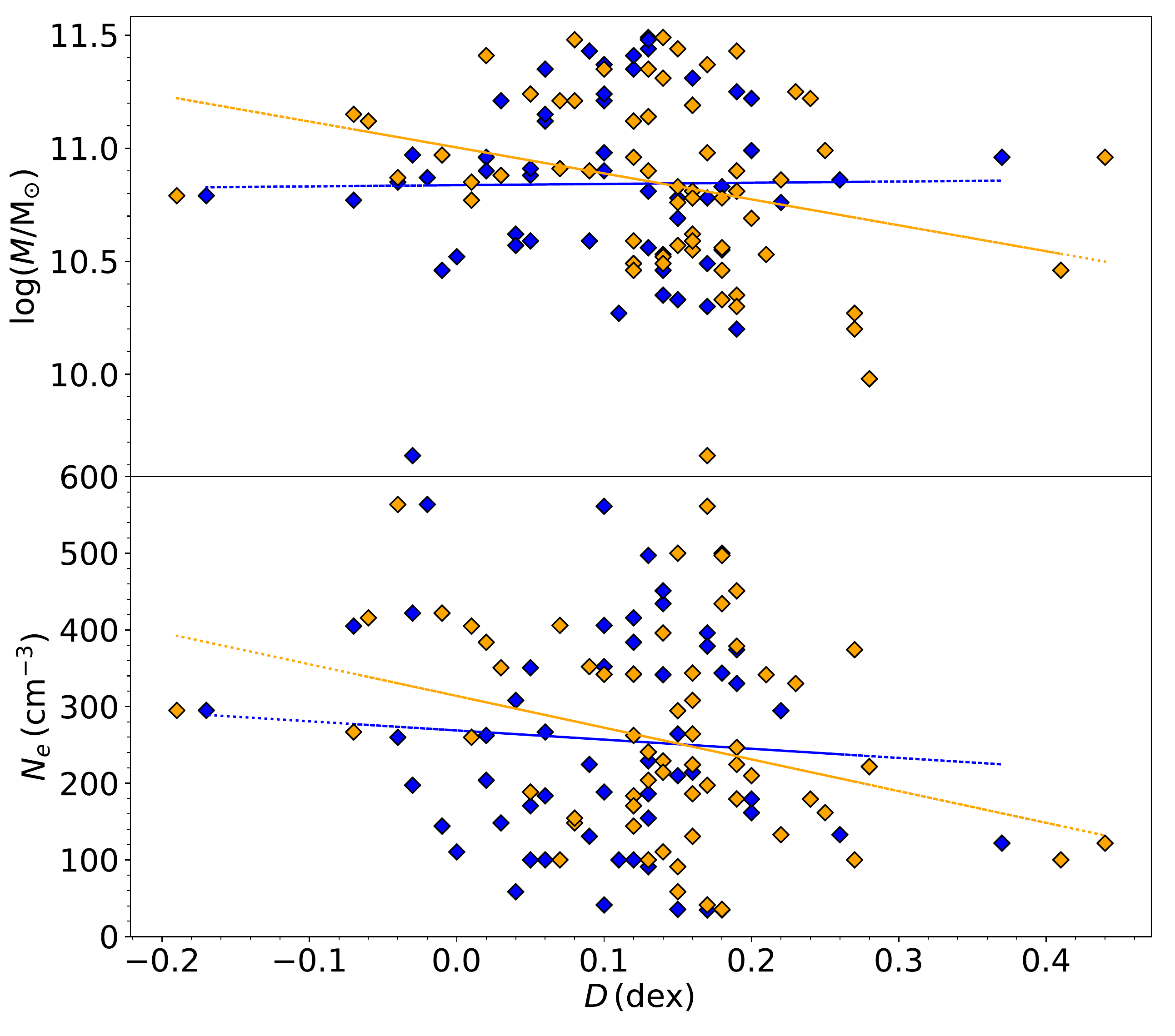}
\includegraphics[width=1.02\columnwidth]{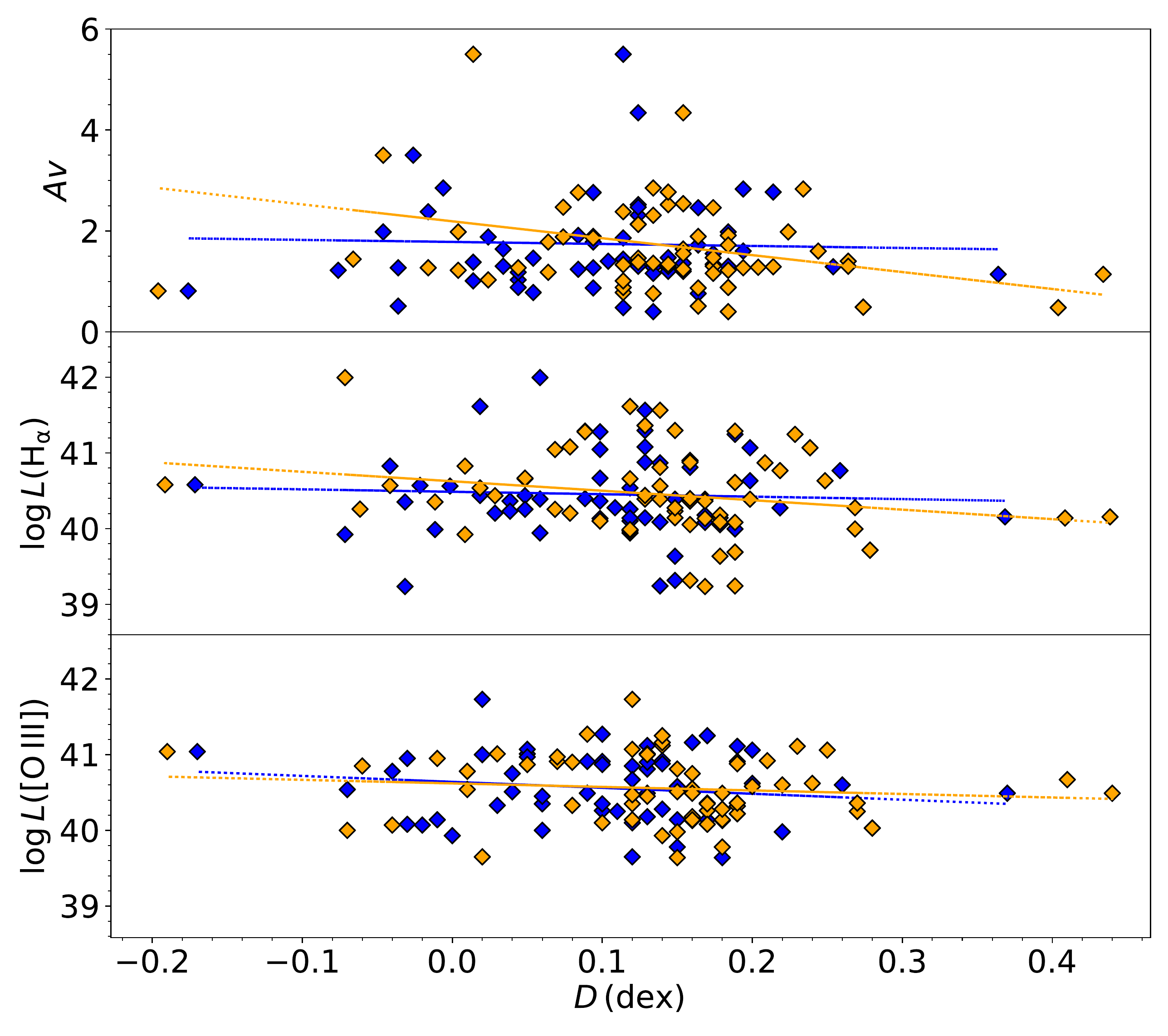}
\vspace*{2mm}
\caption{Left panels: Electron density (in units of cm$^{-3}$) for each AGN NLR and the logarithm of
the stellar mass of the hosting galaxy versus the difference D (in dex, Eq.~\ref{newdf})
between the O/H abundance of the AGN (derived from \citealt{carvalho19} calibration -- Eq.~\ref{calib1_agn})
and the intersect oxygen abundances derived
from the radial abundance gradients -- blue symbols for the HII O3N2 calibration and orange symbols for the N2 calibration of PMC09. 
Rigth panels:
As in the left panels but for the logarithm of the luminosity of [\ion{O}{iii}]$\lambda$5007 and H$\alpha$ as well
 as the extinction $A_{\rm v}$ (in units of magnitude) of the AGNs. Blue and orange lines represent a linear regression  to the corresponding points. The linear regression coefficients as well as the Pearson Correlation Coefficients (R) and p-values for the blue points are listed in Table~\ref{tablean1}, while those for the orange points are listed in Table~\ref{tablean2}.}
\label{fan1}
\end{figure*}

In order to investigate if there is any correlation between the oxygen discrepancy $D$, derived by using  Eq.~\ref{newdf}, and 
relevant physical properties of the AGN sample, in Fig.~\ref{fan1} we plot the electron density of the AGN NLR, luminosity of H$\alpha$ and the
extinction coefficient $A_{\rm v}$ obtained within the inner 2.5 arcsec versus $D$. In addition, the luminosity of [\ion{O}{iii}]$\lambda$5007 and stellar masses derived from SDSS-III data \citep{Thomas+13} were also 
plotted versus $D$ in the second column of Fig.~\ref{fan1}. The blue  and orange symbols represent which calibration
was considered in the radial extrapolation: blue for the $O3N2$ calibration and orange for the $N2$ calibration of PMC09. 
A linear regression to the points in each
plot of Fig.~\ref{fan1} was performed. The best fit coefficients, the Pearson Correlation Coefficient  (R) and the $p$ value
are listed in Tables~\ref{tablean1} and ~\ref{tablean2}. The  O/H abundances  in the AGNs  were 
those obtained through \citet{carvalho19} because this calibration considers a wider
range of nebular parameters than that of \citet{SB98}. 
Based on the linear fits, $R$ and $p$ values  and from the analysis of the plots in Fig.~\ref{fan1}, we can state that for the $O3N2$ calibration (represented by blue symbols),  
there is no correlation between the AGN and galaxy properties and $D$. The $p$ value confirms that R is not significant (considering the level of significance as $p\leq$0.05). On the other hand, for the $N2$ calibration (represented by orange symbols), there is evidence of a mild inverse correlation between the following properties and $D$: electron density, stellar mass and extinction $A_{\rm v}$. The cause of these inverse correlations is not clear. 

It is worthwhile to stress again that, the oxygen abundance estimations via strong line methods for \ion{H}{ii} regions and AGNs can differ from each other up to $\sim 0.8$ dex (e.g. \citealt{Kewley-ellison+08, dors20b}).
Thus, to confirm the non-existence of correlation between
AGN nebular parameters and $D$ found above it is necessary to estimate the  O/H gradients and AGN abundances
based on direct estimations of the electron temperature, i.e. by using the $T_{\rm e}$-method (the most reliable method), which is not possible considering the data  in this paper.


\begin{table}
 \caption{Coefficients  of the linear regression  to the blue symbols (O3N2 HII regions O/H calibration)
 shown Fig.~\ref{fan1} and the corresponding Pearson Correlation Coefficient (R) and the $p$-value.}
 \begin{tabular}{lcccc}

\hline    
                          &       a        &  b                 & R        &$p$\\
\hline             
$N_{\rm e}$ (cm$^{-3}$)   &  $-$119$\pm194.1$    & 268.8$\pm27.19$    & $-$0.08         & 0.54 \\
log($M/{\rm M_{\odot}}$)  &  $0.05\pm0.57$ & 10.84$\pm0.08$    & $-$0.01       & 0.93\\
log$L$([\ion{O}{iii}])    & $-$0.79$\pm0.63$  & 40.64$\pm0.09$    &  $-$0.16.     & 0.22\\
log$L$(H$\alpha$)         &  $-$0.32$\pm0.83$  & 40.49$\pm0.12$    &  $-$0.05      & 0.70\\
$A_{\rm v}$               &  $-$0.40$\pm1.58$  &  1.78$\pm0.22$    & $-$0.03        & 0.82\\
\hline
\end{tabular}
\label{tablean1}
\end{table}

\begin{table}
 \caption{Coefficients  of the linear regression  to the orange symbols (N2 HII regions O/H calibration)
 shown Fig.~\ref{fan1}, the corresponding Pearson Correlation Coefficient (R) and the $p$-value.}
 \begin{tabular}{lcccc}
\hline    
                          &       a        &  b                 & R        &$p$\\
\hline             
$N_{\rm e}$ (cm$^{-3}$)   &  $-$413.7$\pm161.2$    & 313.8$\pm27.87$    & $-$0.32         & 0.01 \\
log($M/{\rm M_{\odot}}$)  &  $-$1.15$\pm0.48$ & 11$\pm0.08$    & $-$0.30       & 0.02\\
log$L$([\ion{O}{iii}])    & $-$0.47$\pm0.56$  & 40.62$\pm0.10$    &  $-$0.11.     & 0.41\\
log$L$(H$\alpha$)         &  $-$1.25$\pm0.71$  & 40.63$\pm0.12$    &  $-$0.22      & 0.09\\
$A_{\rm v}$               &  $-$3.35$\pm1.31$  &  2.21$\pm0.23$    & $-$0.32        & 0.01\\
\hline
\end{tabular}
\label{tablean2}
\end{table}

\section{SUMMARY}
\label{conc}

We derived the metallicities of MaNGA AGN NLRs (traced by the O/H abundance) and the 
radial gradients of oxygen abundance along the disk for 98 Seyfert~2 and 10 Seyfert~1 host galaxies using MaNGA-SDSS-IV data cubes. The metallicities of the AGNs  and in the disk of \ion{H}{ii} regions  were obtained using calibrations based on strong emission lines proposed in the literature. We derived for
most galaxies  clear O/H gradients, with the O/H abundance ratio decreasing as the galactocentric distance increases. This characteristic is commonly found in the disk of spiral galaxies, which suggest that most spiral galaxies are formed according to the inside-out scenario.
The oxygen abundances derived through emission lines of the AGNs and based on two distinct 
calibrations are lower by an average value $<D>$ of 0.16-0.30 dex (depending on the calibration assumed)
than the extrapolated oxygen abundances to the central parts  derived
from the radial abundance gradients. 
We  suggest that the difference $<D>$ can be due to 
the accretion of metal-poor gas to the AGN host -- probably via the capture of a gas-rich dwarf galaxy, that builds up 
a reservoir of molecular and/or neutral gas which will then feed the SMBH. This gas will then trigger the nuclear acitivity via its capture by the nuclear 
supermassive black hole (SMBH).
We investigated correlations between  $D$
and the electron density ($N_{\rm e}$), [\ion{O}{iii}]$\lambda$5007 and H$\alpha$ luminosities, extinction coefficient 
($A_{V})$ of the AGN as well as the stellar mass ($M_{*}$)of the hosting galaxy. We did not find any significant correlation between the  aforementioned  properties and $D$ when the oxygen gradients are derived from $O3N2$ index.
Otherwise, there is evidence of an  inverse correlation between the $D$ and $N_{\rm e}$, $M_{*}$  and  $A_{\rm v}$ when
the $N2$ index is used.
The origin of inconsistency observed here, probably, it is due to the use of different metallicity indicators  to derive the radial gradients. To confirm the derived correlations, further investigation with oxygen radial gradients and AGN estimations based on direct determination of the electron temperature, i.e. by using the $T_{\rm e}$-method, is required.



\section*{Acknowledgements}
JCN and OLD thank  Fundação de Amparo à Pesquisa do Estado de São Paulo (FAPESP, process: 2019/14050-6). RR thanks to Conselho Nacional de Desenvolvimento Cient\'{i}fico e
Tecnol\'ogico  ( CNPq, Proj. 311223/2020-6,  304927/2017-1 and
400352/2016-8), Funda\c{c}\~ao de amparo 'a pesquisa do Rio Grande do
Sul (FAPERGS, Proj. 16/2551-0000251-7 and 19/1750-2),
Coordena\c{c}\~ao de Aperfei\c{c}oamento de Pessoal de N\'{i}vel
Superior (CAPES, Proj. 0001). RAR thanks to CNPq for partial financial support. TSB and SR acknowledge the support of the Brazilian funding agencies FAPERGS and CNPq. We would like to thank the support of the Instituto Nacional de Ci\^encia e Tecnologia (INCT) e-Universe project

Funding for the Sloan Digital Sky Survey IV has been provided by the Alfred P. Sloan Foundation, the U.S. Department of Energy Office of Science, and the Participating Institutions. SDSS acknowledges support and resources from the Center for High-Performance Computing at the University of Utah. The SDSS web site is www.sdss.org.

SDSS is managed by the Astrophysical Research Consortium for the Participating Institutions of the SDSS Collaboration including the Brazilian Participation Group, the Carnegie Institution for Science, Carnegie Mellon University, the Chilean Participation Group, the French Participation Group, Harvard-Smithsonian Center for Astrophysics, Instituto de Astrof\'isica de Canarias, The Johns Hopkins University, Kavli Institute for the Physics and Mathematics of the Universe (IPMU) / University of Tokyo, Lawrence Berkeley National Laboratory, Leibniz Institut f\"ur Astrophysik Potsdam (AIP), Max-Planck-Institut f\"ur Astronomie (MPIA Heidelberg), Max-Planck-Institut f\"ur Astrophysik (MPA Garching), Max-Planck-Institut f\"ur Extraterrestrische Physik (MPE), National Astronomical Observatories of China, New Mexico State University, New York University, University of Notre Dame, Observat\'orio Nacional / MCTI, The Ohio State University, Pennsylvania State University, Shanghai Astronomical Observatory, United Kingdom Participation Group, Universidad Nacional Aut\'onoma de M\'exico, University of Arizona, University of Colorado Boulder, University of Oxford, University of Portsmouth, University of Utah, University of Virginia, University of Washington, University of Wisconsin, Vanderbilt University, and Yale University.

\section*{Data Availability}

 The data underlying this article are available under SDSS collaboration rules, and the by products will be shared on reasonable request to the corresponding author.







\appendix

\section{AGN maps}

We present all the maps and gradients of the chemical abundances obtained from the calibrations for the AGNs and the controls, as well as the extrapolations in Figs. A1 - A75.

\includepdf[pages=-]{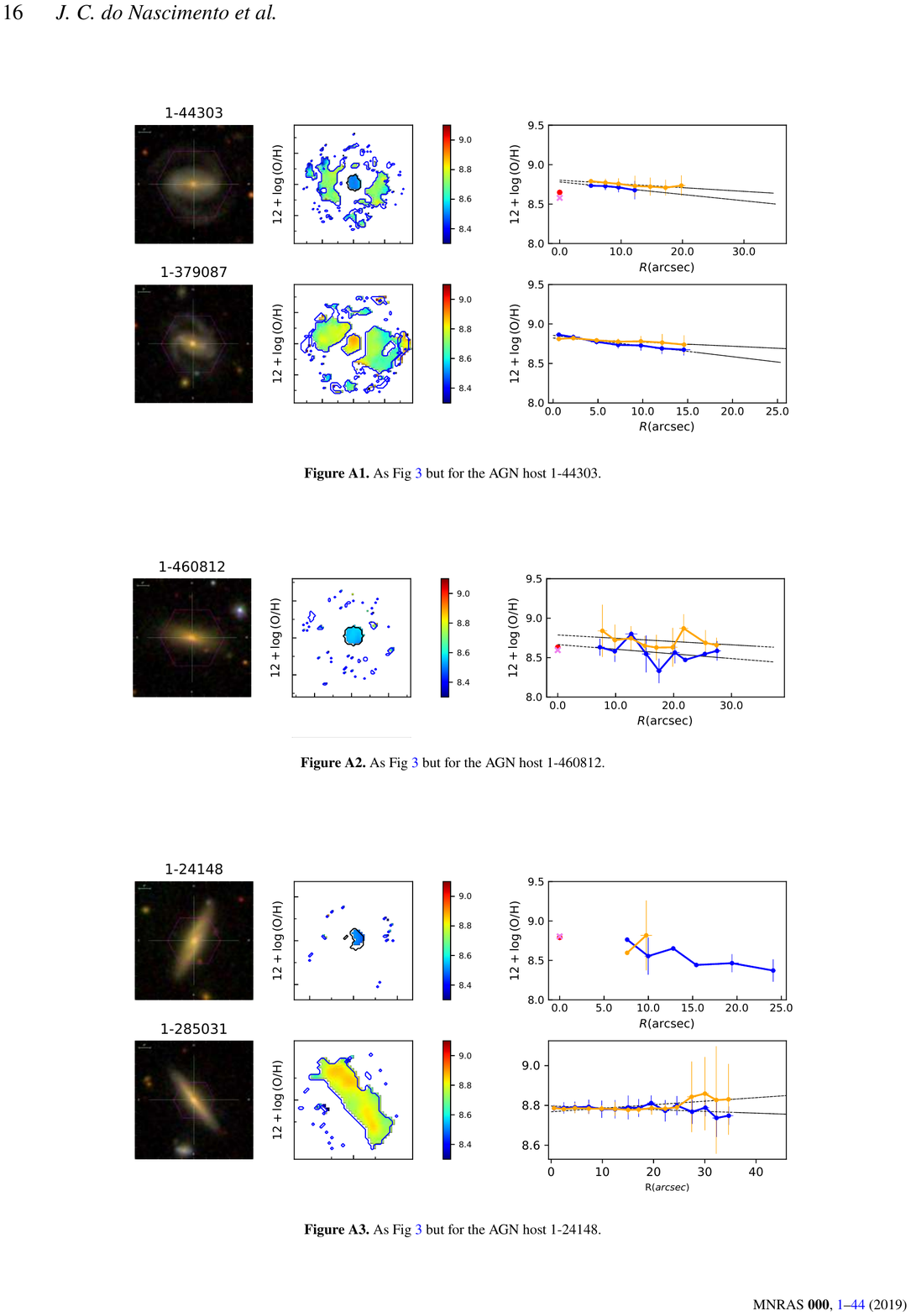}


\bsp	
\label{lastpage}
\end{document}